\documentstyle[aps,eqsecnum,prb,twocolumn,epsf]{revtex}

\begin{document}
\draft
\wideabs{
\title{Classification of Disordered Phases of Quantum Hall Edge States}
\author{Joel E. Moore and Xiao-Gang Wen}
\address{Department of Physics, Massachusetts Institute of Technology,
Cambridge, MA 02139}
\date{October 20, 1997}
\maketitle
\begin{abstract}
The effects of impurity scattering on a general Abelian fractional quantum
Hall (FQH) edge state are analyzed within the chiral-Luttinger-liquid model
of low-energy edge dynamics.  
We find that some disordered edges
can have several different phases characterized by
different symmetries.  The stable impurity edge phases
are in general more symmetric than the original clean system
and demonstrate the phenomenon of dynamical symmetry restoration
at low energies and long length scales.
The phase transitions between different disordered phases
are characterized by broken symmetries and obey Landau's symmetry breaking
principle for continuous phase transitions.
Phase diagrams for various edges are found using a new
system of coordinates for the interactions between modes
in a quantum Hall edge.  The temperature dependence of
tunneling through a point contact is calculated and is found to be able
to distinguish different impurity edge phases of the same
FQH state.
\end{abstract}

}
\section{Introduction}

It was realized soon after the discovery of the integer quantum Hall effect
(QHE) that interesting phenomena occur at the one-dimensional boundary of
a two-dimensional electron gas.~\cite{klitzing}  In a strong applied magnetic field, the bulk
electron gas forms an incompressible quantum liquid~\cite{laughlin}
at certain filling
factors $\nu$ which are found experimentally to be either integers or
simple fractions.  The only gapless excitations at these filling factors
are along the edge of the liquid and as a result current flow is
confined to the edge.~\cite{halperin1}
The low-energy excitation spectrum at the edge
is accessible to tunneling~\cite{milliken} and
magnetoplasma~\cite{ashoori} experiments and in
principle allows the structure of complicated FQH liquids
to be probed because of the connection between the internal topological
orders of the bulk electron gas~\cite{rev,zhang}
and the ``chiral Luttinger liquid'' theory of the edge.~\cite{wen1}

The properties of disordered quantum Hall edges are important for a number
of reasons.  The edge is described by a chiral Luttinger liquid ($\chi$LL)
theory similar to the ordinary
Luttinger liquid,~\cite{haldane1} the
generic state of a one-dimensional interacting electron gas, which
is known to be sensitive to impurities.  In
fact, the difficulty involved in fabricating sufficiently clean
and conducting one-dimensional electron gases has led to interest in quantum
Hall edges as an ideal one-dimensional system.~\cite{beenakker}
The quantum Hall edge
can be impervious to disorder, as in the $\nu = 1$ state, which has a single
branch of low-energy excitations propagating in
one direction and hence remains conducting when random impurities are added.

The effects of disorder on a nonchiral Hall edge
(one with excitations moving in both directions) are more complex.  The
$\chi$LL theory of a clean edge with $n$ condensates is characterized by
two matrices $K$ and $V$ and a charge vector {\bf t}: the $K$ matrix
describes the
topological orders of the bulk state, such as the relative statistics
of quasiparticles, and the $V$ matrix gives the edge Hamiltonian and
related properties such as velocities.  $K$ is taken to be the same for
all edges of the same quantum Hall state, while the values in $V$ are
nonuniversal and expected to vary for different experiments.
The conductance of a maximally
chiral edge (all modes propagating in the same direction) is independent of
$V$ and hence universal.  For clean nonchiral edges, the conductance
calculated using the Kubo formula depends on $V$ and thus appears to be
nonuniversal, contradicting experiment.  The conductance is bounded below
by the quantized value $\nu e^2 / h$ associated with the bulk filling factor
$\nu$ but only has this value for a subset of the allowed $V$ matrices.
However, in real samples, the different modes at the edge equilibrate
and the conductance takes the quantized value.
Kane, Fisher, and Polchinski (KFP) argued that for the $\nu = 2/3$ edge
this equilibration is caused by scattering from random
impurities.~\cite{kfp}

Haldane has argued that an additional term must
be included in the Kubo conductance formula to account for
contributions from the bulk QHE liquid.~\cite{haldane3}  
With this term the conductance
is always fixed at the quantized value $\nu e^2 / h$, even for a nonchiral edge.
This argument involves some subtle questions which have been partly explored
in later work;~\cite{kane3} here we will just mention that disorder-driven
instabilities can affect other measurable properties besides the conductance,
such as tunneling behavior through a point contact.  Thus such instabilities
are relevant to experiments whether or not the original use of the Kubo
formula is correct.  One result of our work is that for some edges
measurably different phases can occur even when the conductance, calculated
with or without the additional term, is fixed at the quantized value.
 
The equilibration of different edge modes is also an important process
in the integer quantum Hall effect with $\nu = n$, since nonideal contacts
will populate
the $n$ different edge channels at different chemical
potentials.  Inter-channel electron scattering can equilibrate the modes,
as predicted by B\"uttiker~\cite{buttiker} and
demonstrated in several innovative
experiments.~\cite{alphenaar,kouwenhoven,takagaki}  Experiments show
equilibration in the IQHE on a length scale
$\ell_e \sim 40\,\mu {\rm m}.$  Nonchiral FQH edges
differ in that even
ideal contacts do not give rise to an equilibrated edge.  Some interactions
capable of transferring charge between channels, such as backscattering by
random impurities and/or electron-phonon interactions, are necessary for 
edge equilibration.~\cite{kane3} In real samples the different branches of FQH 
edges are always close to each other and the different
branches are always in 
equilibrium, which leads to a quantized Hall conductance.

The $\nu=2/3$ edge contains one branch propagating in each direction.
KFP showed that when impurity scattering is relevant, it can drive the velocity
matrix $V$ to one of the subset of possible values which give
conductance quantization.  This subset of ``charge-unmixed'' $V$ matrices
has a simple physical
property: one eigenmode of edge fluctuation is charged and the others are
neutral, with no interaction between the charged and neutral eigenmodes.
Impurity scattering is not relevant for every velocity
matrix, so conductance may not always be quantized.  But for all
velocity matrices sufficiently close
to a charge-unmixed matrix, weak impurity scattering drives the edge state
to the fixed point, where the velocity matrix is charge-unmixed and the
conductance is quantized.

The $V$ matrices for current
experimental setups may all fall within the basin of attraction of the
KFP fixed point.
A velocity matrix is close to a charge-unmixed matrix if the
interaction between
charged modes has much higher energy than the interaction between a charged
mode and a neutral mode.
The Coulomb interaction between charges in an experimental QHE setup is
typically only screened at a distance of many magnetic lengths
$\ell = \sqrt{\hbar c / e B}$ from the edge, so that the
charge-charge interaction may indeed
be much larger than the residual interaction
between a charged mode and a neutral one.~\cite{longrange}

The flow to the impurity fixed point is especially interesting
from the point of view of symmetries of the $\chi$LL action.
The $K$ matrix of the state $\nu = n / (2 n - 1)$ has a hidden $SU(n)$
symmetry, as first pointed out by Read.~\cite{read}
A generic $V$ matrix breaks the symmetry of the $K$ matrix, but
precisely at the impurity fixed point, the velocity matrix does have
all the symmetries of $K$.~\cite{kane}
The impurity scattering thus acts to
{\it increase} the symmetry of the edge.  In this paper we find that in
some quantum Hall edges the impurity fixed points have some but not all
of the symmetries of the $K$ matrix.  In this case different fixed points are 
related by symmetry transformations, 
in the same way as the spin-up and spin-down
fixed points of the Ising ferromagnet in zero external field
are different but are related by a symmetry of the
starting Hamiltonian.  This new type of broken symmetry
is discussed in Section IV.
Edge states differ from ferromagnets in that symmetry is ``spontaneously
restored'' rather than spontaneously broken: a starting Hamiltonian
of low symmetry is driven by impurity scattering to a more
symmetric fixed point.

In this article we analyze the effects of impurity scattering on a general
nonchiral quantum Hall edge.  Not all edge states have
a potentially relevant impurity scattering operator,
which is required for impurity scattering to cause edge
equilibration.  For example,
it is shown in Section III that
the only nonchiral level two states in the Haldane-Halperin
hierarchy~\cite{haldane2,halperin2}
with a possibly relevant impurity scattering operator
are the principal hierarchy states with filling factor
$\nu = 2/(2 p + 1)$, $p$ an odd integer.  The other level two states,
such as $\nu = 4/5$, have an equilibration
length from impurity scattering which diverges at low temperature.~\cite{kane3}
Edges with no relevant operators may still equilibrate by another
process such as inelastic scattering from phonons.
In order to determine whether a given state
flows to an impurity fixed point, it is necessary to consider all the
potentially relevant scattering operators in the chiral Luttinger
liquid theory.

Section II defines a function $K(m)$ whose absolute
value is twice the minimum scaling dimension of the vertex operator
$O_{\bf m} = \exp(i \sum_j m_j \phi_j)$,
where the $m_j$ are integers and $\phi_j$ are the
fields of the $\chi$LL theory.  For impurity scattering $O_{\bf m}$ has
charge zero and $K({\bf m})$ is an even integer.
Impurity scattering operators with $|K({\bf m})| \geq 4$ are never
relevant.  Equilibration by impurity scattering depends on having scattering
operators with $|K({\bf m})| = 2$
to drive the velocity matrix to a charge-unmixed matrix.  Haldane has
argued that a neutral operator $O_{\bf m}$ with $K({\bf m}) = 0$ drives
a topological instability which
removes a pair of oppositely directed neutral modes from the low-energy
theory.~\cite{haldane3}  Edges with no neutral $K({\bf m}) = 0$ operators are
``$T$-stable'' and it is conjectured that only $T$-stable states are
seen experimentally.  All the examples studied in this paper are
$T$-stable, but the methods introduced do not depend on $T$-stability.

$T$-stability is useful because there are only finitely many classes
of $T$-stable states with neutral modes in both directions, as we now
explain.
In Section II we introduce a
system of coordinates for the $V$ matrix
which simplifies the treatment of states with
several condensates.  The dimension of the coordinate space is less than the
number of free parameters in $V$. For states with all neutral
modes traveling opposite to the charge mode, the charge-unmixed subset is
a single point in these coordinates and a
small number of $|K({\bf m})| = 2$ operators
are relevant in a region around this
point, which is the only impurity fixed point.
This class includes
the $\nu = n/(2 n - 1)$ states.  For dim $K = 3$ there are also states
in which the charge-unmixed matrices form a line in our coordinate space
and there are infinitely
many $|K({\bf m})| = 2$ operators.  These states have many
impurity fixed points.
With dim $K = 4$ the
charge-unmixed states can form a plane or a point.
For dim $K = 5$ there are no principal hierarchy states and for dim $K > 5$
no states at all which are $T$-stable and have neutral modes moving in both
directions, as a consequence of a deep theorem on integral quadratic
forms.~\cite{haldane3,watson}


Section III studies $T$-stable hierarchical
quantum Hall states and finds that several classes of such states exhibit
``$V$ stability'': every $V$ matrix sufficiently close to a charge-unmixed
matrix is driven to a charge-unmixed matrix by weak impurity scattering.
In particular, every principal hierarchy state with two or three condensates
is shown to have this property.
Some of the three-species states have infinitely many possibly
relevant operators which lead to many impurity fixed points with
different charge-unmixed $V$ matrices.  For $V$ stable states,
impurity scattering can explain the edge equilibration and robust
quantization seen experimentally.

The rest of this article is organized as follows.
Section II puts the model of a general Abelian quantum Hall edge with
impurities
into a form which isolates the dependence of scaling dimensions
of various operators on the velocity matrix.  This is convenient for the
calculation of phase diagrams for particular edges in Section
III.  Section III applies the formalism to states in the
hierarchy containing several species of quasiparticles and finds
new behaviors associated with the existence of a large number of
possibly relevant operators.
Two classes of edges with high symmetry are studied in
some detail: the $SU(n) \times U(1)$ edge solved exactly by Kane and
Fisher~\cite{kane} and the ``Fibonacci'' edge, in which the sequence
$a_{n+1} = a_n + a_{n - 1}$ plays a special role.  All principal hierarchy
states with three condensates are shown to belong to one of
these two classes.  The Fibonacci edge has two types of fixed points which
correspond to different phases with different tunneling conductance
and other measurable properties.  Some four-condensate edges, such as
$\nu = 12/17$, are shown to have three different types of fixed points,
representing three different broken symmetries of the $K$ matrix.

Sections IV-VI contain results obtained using the method developed
in the earlier, more technical sections and require only a general
understanding of the earlier sections.  Section IV explains how
the many fixed points in some edges are related to broken
symmetries of the $K$ matrix.  Section V examines the
experimental consequences of the multiple phases in disordered edges.
The low-temperature scaling behavior of the tunneling conductance
through a point constriction is calculated for all phases of all
$T$-stable principal hierarchy states with filling fractions $\nu > 1/4.$
This experiment is capable of distinguishing different impurity phases
of the same edge state.  In Section VI we summarize our results from
the point of view of general principles of phase transitions.

\section{General properties of the disordered edge}
Edges of quantum Hall systems are described by a chiral Luttinger liquid
($\chi$LL) theory related to the topological orders of the
bulk quantum Hall state.
We introduce the theory for a clean edge and diagonalize it to
obtain scaling dimensions of impurity scattering operators.
The $\chi$LL action in imaginary time for a clean edge of a QH state
characterized by the matrix $K$ contains
$n = {\rm dim}\ K$ bosonic fields $\phi_i$ and has the form~\cite{wen1}
\begin{equation}
\label{sec:action}
S_0 = {1 \over 4 \pi} \int{dx \, d\tau\, [K_{ij} \partial_x \phi_i \partial_\tau \phi_j +
V_{ij} \partial_x \phi_i \partial_x \phi_j]}
\end{equation}
where, as in the rest of this paper, the sum over repeated indices is assumed.
$K$ is a symmetric integer matrix and $V$ a symmetric positive matrix. 
$K$ gives the topological properties of the edge: the types of
quasiparticles and their relative statistics.  $V$, the velocity matrix, is positive
definite so that the Hamiltonian is bounded below.  The charges of
quasiparticles are specified by an integer vector ${\bf t}$ and the filling
factor is $\nu = t_i (K^{-1})_{ij} t_j$.

Scattering by spatially random quenched impurities is described by the action
\begin{equation}
\label{sec:action2}
S_1 = \int{dx \, d\tau \, [\xi(x) e^{i m_j \phi_j} + \xi^*(x)
e^{-i m_j \phi_j}]}
\end{equation}
Here $\xi$ is a complex random variable and
$[\xi(x) \xi^*(x^\prime)] = D \delta(x - 
x^\prime)$, with $D$ the (real) disorder strength.
The integer vector ${\bf m}$ describes how many
of each type of quasiparticle are annihilated or created by the operator
$O_{\bf m} = \exp(i m_j \phi_j)$.  For a real system all charge-neutral
scattering operators $m_j$ are expected to appear, but most of these will be
irrelevant in the RG sense as
discussed in the following.  The condition for charge-neutrality is
$t_i (K^{-1})_{ij} m_j = 0$.
The random variables $\xi_{\bf m}$ for different scattering
operators $O_{\bf m}$ may be
uncorrelated or correlated depending on the nature of the physical impurities causing the scattering.

Now consider the correlation functions of these scattering operators with
respect to the clean action $S_0$.  For integer vectors ${\bf m}$, define the 
function $K({\bf m}) \equiv m_i (K^{-1})_{ij} m_j$.  $K({\bf m})$ 
governs the topological part of
the correlation function of the scattering operator $O_{\bf m}$
as follows: the correlation function is,
ignoring cutoffs,
\begin{eqnarray}
&G(x,\tau) = \langle e^{i m_j \phi_j(x,\tau)} e^{-i m_j \phi_j(0,0)} \rangle \nonumber
\\
&\propto
(\prod_{k=1}^{n^+} (x + i v^+_k \tau)^{-\alpha_k}) (\prod_{k=1}^{n^-}
(x - i v^-_k \tau)^{-\beta_k}).
\label{sec:exponents}
\end{eqnarray}
Here $n^+$ and $n^-$ are the numbers of positive and negative eigenvalues
of $K$, and $v^\pm_k, \alpha_k, \beta_k$ are nonnegative real numbers which
depend on $V$ and $K$.  However, $\sum_{k=1}^{n^+} \alpha_k - \sum_{k=1}^{n^-}
\beta_k = K({\bf m})$ independent of $V$.  Setting all velocities $v^{\pm}_k = 1$
and introducing $z = x + i \tau$,
\begin{equation}
\langle e^{i m_j \phi_j(x,\tau)} e^{-i m_j \phi_j(0,0)} \rangle \propto
{1 \over z^{K({\bf m})}} {1 \over |z|^{2 \Delta({\bf m}) - K({\bf m})}}
\end{equation}
with $K({\bf m})$ assumed positive.  $\Delta({\bf m}) =
(\sum_{k=1}^{n^+} \alpha_k + \sum_{k=1}^{n^-} \beta_k)/2$ is the scaling
dimension of the operator $\exp(i m_j \phi_j)$.

The scaling dimensions of the various operators in the theory
are functions of $V$, an $n \times n$ matrix.  Much of the physics of
a disordered edge
depends on $V$ only through the scaling dimensions of various operators.
The conductance in units of $e^2 / h$ is given by
twice the scaling dimension $\Delta({\bf t})$ of the charge operator as
a consequence of the Kubo formula.~\cite{kfp}
This is the conductance measured with ideal contacts; a kinetic theory
model for nonideal tunnel-junction contacts gives a different nonuniversal
value.~\cite{kane3}  It remains true that edge equilibrium is required
for the universal value of conductance
$\nu e^2 / h$ to be observed, and a necessary condition is
$2 \Delta({\bf t}) = \nu$.

The scaling dimension of an operator determines whether that operator is
relevant in the RG sense when added to the clean action $S_0$.
The operator is relevant with a uniform coefficient when $\Delta({\bf m}) < 2$,
relevant with a spatially random coefficient when $\Delta({\bf m}) < 3/2$,
and relevant at a point (with a $\delta$-function coefficient) when
$\Delta({\bf m}) < 1$.  For the random case this follows from the leading-order
RG flow equation for disorder strength $D$,~\cite{giamarchi}
\begin{equation}
{dD \over dl} = (3 - 2 \Delta) D.
\end{equation}
It is thus useful to write $V$ in a way which isolates
the parts of $V$ which affect $\Delta({\bf m})$ so that scaling dimensions
depend on as few parameters as possible.

Equation \ref{sec:exponents} is obtained by simultaneously diagonalizing
$K$ and $V$ by a basis change $\phi_i = M_{ij}\tilde \phi_j$.  Suppose
$M_1$ brings
$K$ to the pseudo-identity $I(n^{+},n^{-})$, i.e.,
\begin{equation}
M_1^{\rm T} K M_1 = I_{n^{+},n^{-}} = \left(\matrix{
I_{n^{+}}&0\cr
0&-I_{n^{-}}}\right).
\end{equation}
Basis changes preserve the number of positive and negative eigenvalues of a matrix (``Sylvester's
law of inertia'').  Now consider another basis change $M_2$ which will
diagonalize $V$ without affecting
the pseudo-identity: $M_2 \in SO(n^{+},n^{-}) \Rightarrow M_2^{\rm T} I(n^{+},n^{-}) M_2 = I(n^{+},n^{-})$,
introducing the proper pseudo-orthogonal group $SO(m,n)$.  The real positive symmetric matrix
$V^{\prime} \equiv M_1^{\rm T} V M_1$ can be
written as $(M_2^{-1})^{\rm T} V_D M_2^{-1}$ for
some diagonal matrix $V_D$ and some $M_2 \in SO(n^{+},n^{-})$.
The entries in $V_D$ are all positive and
are the $v^{\pm}_k$ from (\ref{sec:exponents}), with $(v^{+},v^{-})$
corresponding to
(positive, negative) eigenvalues of $K$.

Since $V_D$ and $I(n^{+},n^{-})$ are diagonal, the correlation functions
in the basis ${\bf \tilde\phi} =
(M_1 M_2)^{-1} {\bf \phi}$ are trivial:
\begin{eqnarray}
\langle e^{i \tilde \phi_j(x,\tau)} e^{- i \tilde \phi_j(0,0)} \rangle &=&
e^{\langle \tilde \phi_j(x,\tau) \tilde \phi_j(0,0) \rangle - \langle \tilde \phi_j(0,0) \tilde
\phi_j(0,0)\rangle} \nonumber \\
&\propto&
{1 \over x \pm i v_j \tau}
\end{eqnarray} 
where the sign depends on whether $\tilde \phi_j$ appears with $-1$ or $+1$ in $I(n^{+},n^{-})$.
Going back to the original fields ${\bf \phi}$, we obtain
\begin{eqnarray}
\label{sec:eqf}
K^{-1} &=& M_1 M_2 I_{n^{+},n^{-}} M_2^{\rm T} M_1^{\rm T}
= M_1 I_{n^{+},n^{-}} M_1^{\rm T}, \\
V_D &=& M_2^{\rm T} M_1^{\rm T} V M_1 M_2.
\end{eqnarray}
Let us define a matrix $\Delta$ through
\begin{eqnarray}
I &=& M_2^{\rm T} M_1^{\rm T} (2 \Delta)^{-1} M_1 M_2 \nonumber \\
\Rightarrow \quad 2 \Delta &=& M_1 M_2 M_2^{\rm T} M_1^{\rm T}.
\label{sec:eql}
\end{eqnarray}
The positive definite matrix $\Delta$ gives the scaling
dimension of the operator $O_{\bf m}$:
$\Delta({\bf m}) = m_i \Delta_{ij} m_j$.  Note that under the basis change
$\phi_i = M_{ij} {\tilde \phi}_j$, the vector ${\bf m}$ transforms to
preserve ${\tilde m}_i {\tilde \phi}_i = m_i \phi_i =
m_i M_{ij} {\tilde \phi}_j,$ so ${\bf \tilde m} = M^{\rm T} {\bf m}$.  Thus
the functions $K({\bf m})$ and $\Delta({\bf m})$ are basis-invariant. 

The scaling
dimensions are independent of the $n = n^{+} + n^{-}$ velocities in $V_D$,
as expected on physical grounds.  $M_1$ depends only on
$K$, not on $V$, so all possible matrices $\Delta$ for a given edge are obtained as $M_2$ ranges
over $SO(m,n)$ with $M_1$ fixed.  We now introduce a parameterization of $M_2$ in which only
$n^+ n^-$ coordinates affect $\Delta$.  The physical picture is that the scaling dimensions are independent
of the velocities of the eigenmodes and also of the interactions between modes going in the same direction;
the scaling dimensions are only affected by interactions between counterpropagating modes.
Thus of the $n(n+1)/2$ free parameters in $V$, $n$ correspond to velocities of eigenmodes,
$(n^+ (n^+ -1) + n^- (n^- - 1)) / 2$ to same-direction interactions, and $n^+ n^-$ to opposite-direction
interactions.

The study of a nonchiral edge with several branches of excitations is thus
feasible if one is willing to concentrate on edge properties and
renormalization-group flows determined by
the scaling dimensions of various operators.  There are interesting physical
phenomena which are not determined solely by scaling dimensions, such as the
equilibration of velocities of modes moving in the same direction by
interchannel hopping (which does not affect the conductance).  But
the effects of disorder on the commonly measured physical properties
can be obtained from studying only the $n^+ n^-$
parameters of $V$ which affect scaling dimensions, rather than the $n (n+1)/2$
needed for a complete description of the theory.  This is apparent in the
study of an $n=2$ case ($\nu = 2/3$)\cite{kfp}: the velocity matrix
has the form
\begin{equation}
V = \left(\matrix{v_1 & v_{12} \cr
v_{12} & 3 v_2}\right)
\end{equation}
with one branch in each
direction, and the conductance and the structure of the RG flow are
found to depend only on the combination $c = 2 v_{12} (v_1 + v_2)^{-1}
/ \sqrt{3}$.

The separation of $V$ comes about because every element $M$ in
$SO(m,n)$ can be written as a product of a
symmetric positive matrix $B$ and an orthogonal matrix $R$, both of which are in $SO(m,n)$.  This is a
generalization of the familiar decomposition of a Lorentz transformation (an element of $SO(3,1)$)
into a boost (a symmetric positive matrix) and a rotation (an orthogonal
matrix).  For all examples in this paper $m = 1$ or $n = 1$ and this decomposition follows
easily from the parameterization of boost matrices given below.
More details are in Appendix \ref{boostapp}.  Writing $M_2 = B R$,
\begin{equation}
2 \Delta = M_1 M_2 M_2^{\rm T} M_1^{\rm T} = M_1 B R R^{\rm T} B^{\rm T} M_1^{\rm T} = M_1 B^2 M_1^{\rm T}.
\end{equation}
So $\Delta$ is independent of $R$ and depends only on the $n^+ n^-$ parameters in $B$.  $B$ can be written
\begin{equation}
B = \exp\left(\matrix{{\bf 0}& b \cr b^{\rm T} & {\bf 0}}\right)
\end{equation}
for some $n^{+} \times n^{-}$ matrix $b$.

For a maximally chiral edge, the boost part $B$ is just the identity
matrix, so the scaling
dimension of every operator $\exp(i m_j \phi_j)$ is independent of $V$,
and in particular the conductance $\sigma = 2 \Delta(t) = K(t) = \nu$.
For nonchiral edges, nonuniversal values of the conductance are possible
with $2 \Delta(t) \geq \nu$ and equality if and only if the velocity
matrix is charge-unmixed.
This is a special case of the general property $2 \Delta({\bf m}) \geq
|K({\bf m})|$ for all integer vectors ${\bf m}$ (with equality if and
only if the $V_{1j}$ vanish in the basis with ${\bf e}_1 \parallel
{\bf m}$ and $K^{-1}$ diagonal).
Consequently the scattering term $\exp(i m_j \phi_j)$ can only be
relevant if $|K({\bf m})| \leq 3$.  The scattering operator must have bosonic
commutation relations so the three possibilities are $K({\bf m}) = 2,0,-2$.
If a null vector exists with $K({\bf m}) = 0$, the edge is not $T$-stable.
Operators with $|K({\bf m})| = \pm 2$ are necessary if the impurities are
to drive the edge state to a fixed point.  The next step is to calculate
which velocity matrices make the scattering terms $\xi(x) \exp(i m_j \phi_j)
+ \xi^{*}(x) \exp(-i m_j \phi_j)$ relevant.

The possible matrices $\Delta$ for a given edge can be studied simply
by calculating $2 \Delta = M_1 B^2 M_1^{\rm T}$ for all boosts $B$.  For
a two-component edge with one branch propagating in each direction,
there is just one boost parameter.  For a three-component edge, there are
two parameters, and the scaling dimensions of various operators can be
plotted on the plane as functions of these two parameters.
For $SO(1,m)$ a useful parameterization of boosts as a function of
momentum coordinates
$(p_1,\ldots,p_m)$ is~\cite{misner}
\begin{eqnarray}
B_{11} &=& \gamma = \sqrt{1 + p^2}, \quad B_{1i} = B_{i1} = p_{i-1},
\nonumber \\
B_{ij} &=& \delta_{ij} + p_{i-1} p_{j-1} (\gamma - 1) / p^2
\label{sec:boostdef}
\end{eqnarray}
where $2 \leq i,j \leq m + 1$.  It is convenient to work with dimensionless
momentum ${\bf p} = \gamma {\bf v}$ because of the
singularity at $v = c = 1$ in the velocity coordinates.
However, in Section III we mention an advantage of the velocity
coordinates for certain edges.  Permuting indices gives a version
appropriate for $SO(m,1)$.

For a given edge it is now possible to search for all possibly relevant
neutral operators
($|K({\bf m})| = 2$) and then calculate where in the space of
boost parameters each
operator is relevant.  The rest of this section describes a few technical
details needed to carry out this program.  The search for
$|K({\bf m})| = 2$ operators is done on
a computer: there is a finite p-adic test for whether an integer quadratic
form takes the value zero,~\cite{watson} but we know of none to determine all
vectors for which an integer quadratic form takes a particular nonzero value.
When finding phase diagrams in the next section, it will be useful to
consider basis changes
not in $SL(n,Z)$ which bring $K^{-1}$ to diagonal form, so that
the locality condition is no longer that ${\bf m}$ be an integer vector.
The local operators in the new theory are the transforms of integer vectors
in the original theory.  The advantage of such a basis change
$K^{-1} \rightarrow O K^{-1} O^{\rm T}$, ${\bf m}
\rightarrow {O^{\rm T}}^{-1} {\bf m}$ which makes $K^{-1}$ diagonal and
brings the charge vector ${\bf t}$ to the first basis
vector ${\bf e_1}$ is that
some of the boost parameters can be interpreted as the strength of mixing
of the charge mode with neutral modes.  Then the charge-unmixed velocity
matrices will be exactly those with these boost parameters equal to zero.
Table \ref{table1} summarizes the possible parameter spaces for all
nonchiral edges with
four or fewer components.

For each operator with $|K({\bf m})| = 2$, there is some velocity
matrix which gives that operator scaling dimension
$2 \Delta({\bf m}) = 2$: this follows from the possibility of choosing
$M_1$ in (\ref{sec:eqf}-\ref{sec:eql})
to make ${\bf m}$ one of the basis vectors and choosing $M_2$ so that all
parameters rotating
${\bf m}$ into other basis vectors are zero.  The operation of changing
bases distorts the phase diagram nonlinearly but preserves its topology
and produces the same set of possible scaling matrices $\Delta$.
The sign of $K({\bf m})$ will turn out to affect the dimension of the subset of
matrices $V$ which make $\exp{i m_j \phi_j}$ maximally relevant.
The examples of different types of edges listed in Table \ref{table1} are the subject
of the following section.

\section{Structure of disordered edges}

The method described in the previous section greatly simplifies the
analysis of a nonchiral edge of several condensates.  In particular
it allows us to determine in which regions of the space of velocity matrices
$V$ a particular impurity scattering operator $\exp(i m_j \phi_j)$ is
relevant, and hence determine the phase diagram of the edge state.
We find that the edge of a single quantum Hall state can have
different phases, with transitions between phases caused by changes
in $V$.
We also find that only a special class of edge states
(``principal hierarchy states'') have enough
$|K({\bf m})| = 2$ impurity scattering operators to ensure that the
conductivity is driven to the quantized value.  The phase diagrams
for this class of edge states show remarkable symmetries
absent in the phase diagram of a general edge.  In Section IV these
symmetries are shown to reflect broken symmetries of the $K$ matrix.

All the examples are in the hierarchy
of quantum Hall states.~\cite{haldane2,halperin2}
Hierarchical states have tridiagonal
$K$ matrices with all off-diagonal
matrix elements equal to 1 and $K_{11} = l$ an odd integer,
$K_{ii} = n_i$ even for $i = 2,\ldots, {\rm dim}\ K$.  The matrix will
often be given simply by its diagonal elements $(l,n_2,\ldots)$.
The charge vector
is ${\bf t} = (1,0,\ldots,0)$.  The number of modes moving opposite the
direction of the charge mode is equal to the number of negative elements on
the main diagonal.  States with all $|n_i| = 2$ are called principal states
and are the most stable states at each level of the hierarchy.

First we study the edges of all hierarchy states
at second level (dim $K = 2$) and show that the principal hierarchy
states are all similar to the $\nu = 2/3$ state studied
by KFP.  The states which are not principal have no relevant random
operators and are thus unaffected by weak impurity scattering.  In particular,
for these states elastic impurity scattering alone is insufficient to
give edge equilibration at low temperature.

A rich variety of behavior is possible for dim $K = 3$ states, where the two
neutral modes can move in the same direction (opposite the charge mode)
or in opposite directions (cf. Table \ref{table1}).  The principal hierarchy
state of dim $K = n$ with all neutral modes in the same
direction flows to an $SU(n) \times U(1)$ fixed point which is the
only point where conductance is quantized.  The charge mode satisfies a
$U(1)$ Kac-Moody algebra, and the $n - 1$ neutral modes satisfy an
$SU(n)$ Kac-Moody algebra.  The highly symmetric phase diagram for
the $SU(3)$ case is shown not to describe the simplest few
non-principal states.

For the dim $K = 3$ case with neutral modes in
both directions, conductance is quantized along a
line in the phase diagram, and for the principal hierarchy states we find
an infinite number of fixed points along this line corresponding
to the infinite number of possibly relevant random operators.
There are two different types of fixed points which correspond to two
measurably different phases.  A few results on the dim $K = 4$ cases
are also presented.  No principal hierarchy edges
with dim $K > 4$ are topologically stable
except those with all neutral modes in the same direction.~\cite{haldane3}


\subsection{Edges with dim $K = 2$}

The $K$ matrix in the hierarchy basis has the form
\begin{equation}
K = \left(\matrix{l & 1 \cr 1 & n}\right), \quad {\bf t} = (1,0)
\end{equation}
with $l$ odd and $n$ even.
For the state to be nonchiral, $n < 0.$
A quick calculation shows that if ${\bf m} = (m_1, m_2)$ is a
charge-neutral $K({\bf m}) = -2$
operator (there are no charge-neutral $K({\bf m}) = 2$ operators),
${m_1}^2 = -2 / n$ which has the solutions $m_1 = \pm 1$ if $n = -2$ and
no integer solution otherwise.  Hence for principal hierarchy states
($n = -2$) there is one complex-conjugate pair of possibly relevant
operators labeled by ${\bf m} = \pm(1,-2)$, while for other
hierarchy states there are no relevant random operators.

For a dim $K = 2$ state there is a single boost parameter $p$ and
a single value of this parameter that makes $V$ charge-unmixed.  It remains
to show that this value is exactly the value which gives the scattering
operator $\exp(i m_j \phi_j)$ its minimum scaling dimension $\Delta = 1$.
In the basis of ${\bf t} = (1,0)$ and
${\bf m} = (1,-2)$, $K^{-1}$ is diagonal with elements
$(\nu,-2)$ and the scaling dimension matrix is
\begin{eqnarray}
\label{sec:scaledim}
2 \Delta &=& \left(\matrix{\sqrt{\nu} & 0 \cr 0 & \sqrt{2}}\right)
B^2 \left(\matrix{\sqrt{\nu} & 0 \cr 0 & \sqrt{2}}\right) \\
&=&
\left(\matrix{\sqrt{\nu} \! & 0 \cr 0 \! & \sqrt{2}}\right)
\left(\matrix{\sqrt{1 + p^2}\!\!\! & p \!\!\! \cr \! p & \!\!\!\sqrt{1 + p^2}} \right)
\left(\matrix{\sqrt{\nu} \! & 0 \cr 0 \! & \sqrt{2}}\right). \nonumber
\end{eqnarray}
The conductance $2 \Delta(t)$ is $\nu \sqrt{1 + p^2}$ and the scaling
dimension $\Delta({\bf m})$ is $\sqrt{1 + p^2}$.  So $\Delta({\bf m}) = 1$ exactly
at the charge-unmixed point ($p=0$), as required.  The region of attraction of
this fixed point is determined by the equation $\Delta({\bf m}) \leq 3/2$
giving $-\sqrt{5}/2 \leq p \leq \sqrt{5}/2$ for $\nu = 2/3$.

Now we briefly outline the exact solution at the fixed point found by KFP which
also shows the stability of the fixed point under RG transformations.
Let the elementary fields in the basis defined above be the charge mode
$\phi_\rho$ and neutral mode $\phi_\sigma$.  At the fixed point,
\begin{equation}
\label{sec:fixedkv}
K = \left(\matrix{\nu & 0 \cr 0 & -2}\right),\quad V =
\left(\matrix{v_\rho & 0 \cr 0 & 2 v_\sigma}\right).
\end{equation}
The three operators $\partial_x \phi_\sigma,
\exp(i \phi_\sigma), \exp(-i \phi_\sigma)$ all have scaling dimension 1
and satisfy an $SU(2)$ algebra.  The action at the fixed point is
\begin{eqnarray}
S &=& \int_{x,\tau}[{\nu \partial_x \phi_\rho \over 4 \pi}
(i \partial_\tau
+ v_\rho \partial_x) \phi_\rho \nonumber \\
&+& {2 \partial_x \phi_\sigma \over 4 \pi} (-i \partial_\tau
+ v_\sigma \partial_x) \phi_\sigma + (\xi(x) e^{i \phi_\sigma}
+ {\rm h. c.})],
\end{eqnarray}
obtained by substituting the fixed point
$K$ and $V$ into (\ref{sec:action},\ref{sec:action2}).
Now the fixed point action can be written in terms of a two-component
Fermi field by introducing an auxiliary bosonic field $\chi$ which
does not affect physical quantities:
$\psi_1 = \exp[i (\chi + \phi_\sigma)/\sqrt{2}],\ 
\psi_2 = \exp[i (\chi - \phi_\sigma)/\sqrt{2}].$
The clean part of the action is diagonal
in the components while the impurity term becomes a hermitian combination of
raising and lowering operators, $\psi_1^\dagger \psi_2$ and $\psi_2^\dagger
\psi_1$, with random coefficients.  The impurity term
is then eliminated by a local $SU(2)$ gauge transformation which preserves
the clean part of the action.  The clean part of the action is just the
action for free chiral fermions.

When the system is near but not at the fixed point, there is a weak coupling
$V_{\rho \sigma} \partial_x \phi_\rho \partial_x \phi_\sigma$ between
the charged and neutral modes.  The scaling dimension of this term
in the original action is 2 so the operator is
marginal with a uniform coefficient.  However, the
$SU(2)$ rotation of $\partial_x \phi_\sigma$ gives this term a random
coefficient and makes it irrelevant.  According to this
picture, once $V$ falls into the basin of attraction of the fixed point,
i.e., $|p|<=\sqrt{5}/2$ in (\ref{sec:scaledim}), it flows to the fixed
point $p = 0$ with $K$ and $V$ given by (\ref{sec:fixedkv}).
Since the boost part of $V$ is uniquely determined at the fixed point,
many physical properties are uniquely determined, such as the
conductance $\sigma = \nu e^2 / h$.  The same technique of
fermionization followed by a gauge transformation solves the $SU(n) \times
U(1)$ fixed point described below.

\subsection{Three-branch edges with parallel neutral modes}

Such edges have both neutral modes antiparallel to the charge mode
(line 3 of Table \ref{table1}).  There is a single charge-unmixed point in the
boost coordinates of Section II.
In the hierarchy representation such edges have $K$ matrix $(l,-n_1,-n_2)$.
The principal hierarchy edges of this type are $\nu = 3/5$
with $K = (1,-2,-2)$,
$\nu = 3/11$ with $K = (3,-2,-2)$, and so forth.
The principal hierarchy edges with $n$ condensates and all neutral
modes antiparallel to the charge mode have an $SU(n)$ symmetry
($n = {\rm dim}\ K$) in their $K$ matrix $(l,-2,\ldots,-2)$, as
first pointed out by Read.~\cite{read}  The filling fraction
is $\nu = n / (n (l + 1) - 1)$.
Kane and Fisher showed~\cite{kane} that each of these edges
has a fixed point with a charge field $\phi_{\rho}$
of dimension $\nu / 2$ and a set of $n - 1$ dimension 1 neutral fields
${\phi_\sigma}^i$
obeying an $SU(n)$ algebra.  There are
$n-1$ roots of $SU(n)$ which correspond
to the $n-1$ operators $\partial_x {\phi_\sigma}^i$.  Now we obtain the
phase diagram for the $n = 3$ case, which is easily generalized to
$n > 3$.

Any neutral operator $O_{\bf m}$ for these edges has negative
$K({\bf m})$ because all neutral modes travel opposite the direction of charge.
There must be $(n^2 - 1) - (n - 1) = n (n - 1)$
operators with $K({\bf m}) = -2$ in order
to obtain the complete $SU(n)$ algebra (here ${\bf m}$ and $-{\bf m}$
are counted
independently).  For $\nu = 3/5$ this requires 6 such operators which
in the hierarchy basis are labeled by
${\bf m} = \pm (0,1,-2), \pm (1,-2,1), \pm (1,-1,-1).$
Now the technique of Section II can be used to find when these operators
become relevant and thus the region of attraction of the fixed point.  For
this case the procedure is described in detail for the sake of
clarity; for subsequent cases some intermediate steps will be skipped.

The basis $\{(1,0,0)$,$(0,1,-2)$,$(2,-3,0)\}$ 
brings $K^{-1}$ to diagonal form
with elements $(3/5,-2,-6)$. The above six operators 
with $K( {\bf m}) = -2$
become $\pm (0,1,0)$, $\pm (0,1/2,1/2)$, $\pm (0,1/2,-1/2).$
At the fixed point point, $V$ is also diagonal in the new basis 
$\{\phi_\rho, \phi_1, \phi_2\}$, and
$\exp(i \phi_\rho)$ has scaling dimension $\nu / 2 = 3/10$,
$\exp(i \phi_1)$ scaling dimension 1, and
$\exp(i \phi_2)$ scaling dimension 3, so that a neutral operator
$\exp(i m_1 \phi_1 + i m_2 \phi_2)$
has scaling dimension ${m_1}^2 + 3 {m_2}^2.$  Let $D$ be the diagonal
matrix with diagonal elements $(\sqrt{3/5},\sqrt{2},\sqrt{6})$, which
are the square roots of twice the scaling dimensions of the basis fields
at the fixed point.
Using the boost parameters $(p_1,p_2)$, $\gamma = \sqrt{1 + {p_1}^2
+ {p_2}^2}$, to parameterize non-diagonal $V$, we find
the scaling dimension matrix in
the basis $\{\phi_\rho, \phi_1, \phi_2\}$ is
\begin{equation}
\label{sec:scale35}
2 \Delta = D
\left(\matrix{\gamma & p_1 & p_2 \cr
p_1 & 1 + {{p_1}^2 (\gamma - 1) \over {p_1}^2 + {p_2}^2} &
{p_1 p_2 (\gamma - 1) \over {p_1}^2 + {p_2}^2} \cr
p_2 & {p_1 p_2 (\gamma - 1) \over {p_1}^2 + {p_2}^2} &
1 + {{p_2}^2 (\gamma - 1) \over {p_1}^2 + {p_2}^2}}
\right)^2 
D
\end{equation}
{}From this equation it is apparent that for $p_1 = p_2 = 0$ (a diagonal $V$
in the new basis) all six
operators have scaling dimension equal to $1$, and this is the only
charge-unmixed point since if $p_i$ is nonzero the charge mode is partly
mixed with the $i$th neutral mode.
Fig.~\ref{fig1} shows the scaling dimension of the possibly relevant operators
as functions of $(p_1, p_2)$.
The scaling dimension of $(0,1,0)$ is
independent of $p_2$ so its contours are exactly vertical.
Note that such a plot can be drawn without any information about
the fixed point.

The interpretation of RG flows from Fig.~\ref{fig1}
is quite simple.  Each relevant
scattering operator causes the velocity matrix to move to
make the operator maximally relevant ($\Delta = 1$).  If the starting
velocity matrix is near the origin, all three operators are
relevant and drive the velocity matrix to the origin, the only
point at which all three are maximally relevant.  The high symmetry of
the graph reflects the $SU(3)$ symmetry of the fixed point.  General
three-species hierarchy states do not have this symmetry in the
phase diagram and do not have enough $|K({\bf m})| = -2$ operators to determine
a unique fixed point.  For example, the $\nu = 7/11$ state $(1,-2,-4)$
and the $\nu = 7/9$ state $(1,-4,-2)$ both have just one $K({\bf m}) = -2$ operator
which is maximally relevant along a line through the origin.  The phase
diagram is like Fig.~\ref{fig1} with only one line instead of three.  Now the
charge-unmixed point has an $SU(2)$ symmetry rather than an $SU(3)$ symmetry
because only one impurity operator is relevant.  It is not clear that the
system flows to this point in the absence of long-range interactions,
even if it starts near the charge-unmixed point, because other points along
the maximally relevant line are also possible fixed points.

The
$\nu = 15/19$ state $(1,-4,-4)$ has no $|K({\bf m})| = 2$ operators at all so
no stable fixed points result from the addition of weak disorder.
States with no $|K({\bf m})| = 2$ operators are predicted to have
diverging equilibration lengths from impurity
scattering as temperature is lowered since impurity scattering
is never relevant.
For the other type of third level hierarchical states, which have
one neutral mode parallel to the
charge mode, the same basic property is seen: only for principal hierarchy
states are there enough $|K({\bf m})| = 2$ operators for impurity
scattering to determine a discrete set of charge-unmixed fixed points.

\subsection{Three-branch edges with antiparallel neutral modes}

These edges have a line in the phase diagram along which the conductance
is quantized, rather than a point as in the previous cases.
For the principal hierarchy states,
there are infinitely many
$|K({\bf m})| = 2$ operators, and these can be enumerated by a simple linear
recursion relation.
Examples are $\nu = 5/3$ with $K = (1,2,-2)$,
$\nu = 5/7$ with $K = (1,-2,2)$, $\nu = 5/13$ with $K = (3,2,-2)$,
and $\nu = 5/17$ with $K = (3,-2,2)$.  These four edges all have the property
that their $|K({\bf m})| = 2$ operators $O_{\bf m}$ form a (vector)
Fibonacci sequence: ${\bf m}_{n+1} = {\bf m}_n + {\bf m}_{n-1}$.
The reason why these four edges have the same Fibonacci pattern is that
their $\chi$LL theories have, up to a minus sign, the same
neutral sector.  Thus properties which
depend only on neutral operators
are shared by all states $K=(l,2,-2)$ and $K=(l,-2,2)$ independent of $l$.
The familiar scalar Fibonacci sequence $(1, 1, 2, 3, 5, \ldots)$ has
previously
appeared in physics in growth models of phyllotaxis.
Note that the property ${\bf m}_{n+1} = {\bf m}_n + {\bf m}_{n-1}$ is
linear and hence independent of basis.

The $\nu = 5/3$ edge is convenient to study because of
its $SL(3,Z)$ equivalence
to the diagonal $K$ matrix with elements $(1,1,-3)$.  The charge vector
in this basis is ${\bf t} = (1,1,1)$.  This gives the state a natural
interpretation as a $\nu = 1/3$ gas of holes in two filled Landau levels.
Also in this basis $K(m_1,m_2,m_3) = K(m_2,m_1,m_3)$.
The $|K({\bf m})| = 2$ operators in this theory are labeled by
${\bf m} = \pm(1,-1,0)$, $\pm(1,0,3)$, $\pm(2,-1,3)$,
$\pm(3,-1,6)$, $\pm(5,-2,9),\ldots$ plus the same list with first and second
elements exchanged.  The sign of $K({\bf m})$ alternates
between terms in this sequence: $(1,-1,0)$ has $K({\bf m}) = 2$ (as befits hopping
between two rightward-moving modes), $(1,0,-3)$ has $K({\bf m}) = -2$, and so forth.
There is an important difference between $K({\bf m}) = 2$ and $K({\bf m}) = -2$
operators: $K({\bf m}) = 2$ operators are maximally relevant along a line
in the phase diagram,
while $K({\bf m}) = -2$ operators are maximally relevant at a single point.
This happens for the same reason that the charge-unmixed region was a
single point for edges with all neutral modes opposite the charge mode.
In a basis with ${\bf m}$ an eigenvector, if there are no other
eigenvectors with the same direction then every boost involves
${\bf m}$ and affects its scaling dimension.  If there are other eigenvectors
in the same direction, there is a nontrivial linear
space of boosts which do not affect the scaling dimension of $O_{\bf m}$.

The scaling dimension of the first few $|K({\bf m})| = 2$ operators for
$\nu = 5/3$ are plotted in Fig.~\ref{fig2} as functions of boost parameters
$(p_n,p_c)$ according to
\begin{equation}
2\Delta = D^\prime
\left(\matrix{1 + {{p_c}^2 (\gamma - 1) \over {p_c}^2 + {p_n}^2} &
p_c & {p_c p_n (\gamma - 1) \over {p_c}^2 + {p_n}^2} \cr
p_c & \gamma & p_n \cr
{p_c p_n (\gamma - 1) \over {p_c}^2 + {p_n}^2} & p_n &
1 + {{p_n}^2 (\gamma - 1) \over {p_c}^2 + {p_n}^2}}\right)
D^\prime.
\end{equation}
This expression for $\Delta$ is in the basis 
${\bf t}=(1,1,1)$ , ${\bf m_1}= (1,-1,0)$ and ${\bf m_2}= (1,1,6)$
in terms of the original basis.
${\phi_\rho,\phi_1,\phi_2}$
with $\phi_1 = (1,-1,0), \phi_2 = (1,1,6)$ in terms of the original basis
(where $K=(1,1,-3)$).
Let ${\phi_\rho,\phi_1,\phi_2}$ be the three boson fields in the new basis.
The diagonal matrix $D^\prime$ has elements $(\sqrt{5/3},
\sqrt{2}, \sqrt{10})$.  This expression for $\Delta$ is similar to
(\ref{sec:scale35}) for $\nu = 3/5$ with two important differences: the
scaling dimension of $e^{i \phi_2}$ is 5 rather than 3 at the fixed point
($p_n=p_c=0$), 
and the timelike row and column of the boost matrix correspond to
$\phi_1$ rather than
$\phi_\rho$, because now it is one of the neutral modes rather than the
charge mode which has no other modes parallel to it.

In the coordinate space $(p_n,p_c)$ (Fig.~\ref{fig2}),
$K({\bf m}) = -2$ operators are relevant
on compact regions and $K({\bf m}) = 2$ operators on noncompact regions
of the plane.  
The fixed points form lines and isolated points in Fig.~\ref{fig2}, where
one operator with $|K({\bf m})| = 2$
is maximally relevant. For fixed points on the charge-unmixed line $p_c = 0$
(the $x$-axis in Fig.~\ref{fig2}), there are two marginal operators with the opposite
sign of $K({\bf m})$.  The $x$-coordinates of these special points are
found by taking
alternately the rational part and the coefficient of $\sqrt{5}$ in
$((1 + \sqrt{5})/2)^n$.  The theory at each of these fixed points is
similar: in a basis bringing the maximally relevant operator
$\exp(i m_j \phi_j)$ to $\exp(i \phi_1)$, $\phi_2$ can
be chosen so that the marginal operators at the fixed point
are $\exp(i (\phi_1 \pm \phi_2) / 2)$,
and $\exp(i \phi_2)$ has scaling dimension $5$ rather than $3$ in the
$\nu = 3/5$ case.  The scaling dimension of the marginal operators is
then $(\Delta(\phi_1) + \Delta(\phi_2)) / 4 = (1 + 5)/4 = 3/2$ as required.
The marginal operators cannot form an $SU(3)$
multiplet with the maximally relevant operator because their scaling
dimensions are different.  We have not been able
to obtain an exact solution of this fixed point.
Appendix \ref{rgapp}
describes the leading-order
RG flows along the charge-unmixed line between points
$A$ and $B$ and addresses the stability of the two types of fixed points.
The reasons for the periodicity of Fig.~\ref{fig2} are
discussed in Section IV.

Several dim $K = 3$ non-principal edges of this type (antiparallel neutral
modes) were studied, and all were found to have too few
$|K({\bf m})| = 2$ operators for the system to flow to a quantized $\sigma$.  
The four hierarchical states
with $K$ matrices $(1,2,-4), (1,-2,4), (1,-4,2), (1,4,-2)$,
$\nu = (9/5,9/13,9/11,9/7)$ are not $T$-stable and have only one
$|K({\bf m})| = 2$ operator.  The two states with $K$ matrices $(1,4,-4)$
and $(1,-4,4)$, $\nu = (17/13,17/21)$ each have a Fibonacci sequence
of $|K({\bf m})| = 4$ operators
as well as one $K({\bf m}) = 2$ and one $K({\bf m}) = -2$ operator.
The resulting phase diagram for $\nu = 17/13$ is shown in Fig.~\ref{fig3}.
Most velocity matrices near the charge-unmixed line are not affected by
either $|K({\bf m})| = 2$ operator.  If the starting $V$ matrix makes
the $K({\bf m}) = -2$ operator relevant, the system is driven by impurity
scattering to the $(0,0)$
point on the charge-unmixed line.  For starting points with
this operator irrelevant, impurity scattering is insufficient to give
edge equilibration at low temperatures.

Tuning the $V$ matrix in the $\nu = 17/13$ state
in principle allows a transition like the
KFP transition for $\nu = 2/3$ to be observed, even if the system
is always on the charge-unmixed line.  Recall that for $\nu = 2/3$
the system has continuous, nonuniversal scaling dimensions as long as $V$
is not too close to the charge-unmixed point.  A Kosterlitz-Thouless
type transition occurs when $|p| = \sqrt{5}/2$ in (\ref{sec:scaledim}),
and for $|p| <= \sqrt{5}/2$ the system has a universal scaling dimension
matrix.
In the $\nu = 17/13$ state, as $V$ is tuned on the charge-unmixed line
the scaling dimension matrix is continuously variable until one
of the disorder operators becomes relevant; then $V$ is driven to one of the
two fixed points, depending on which operator is relevant.  Unfortunately
the $\nu = 17/13$ state is expected to be quite difficult to observe,
as it is a nonprincipal state with three condensates.

\subsection{Edges with dim $K = 4$}

The edges with all neutral modes opposite the charge mode have
a single charge-unmixed point in the three-dimensional space of boost
parameters, while the other two types of edges
(Table \ref{table1}) have a plane of charge-unmixed points.  This section
studies the
charge-unmixed plane of four-condensate $T$-stable principal
hierarchy states and
finds a pattern with high symmetry and three different types
of fixed points, two of which are exactly solvable.  The
states studied have $K = (l,2,-2,2)$ or $K = (l,-2,2,-2)$,
which were shown by Haldane to be the
only dim $K = 4$
$T$-stable principal hierarchy states with neutral modes
traveling in both directions.~\cite{haldane3}
Examples are $\nu = 12/31$ with $K = (3,-2,2,-2)$ and
$\nu = 12/17$ with $K = (1,2,-2,2)$.

These states behave differently away from the charge-unmixed
plane but have identical structures on the plane, where each state
has two neutral modes traveling in one direction and one neutral mode
traveling in the opposite direction as well as a decoupled charge mode.
For definiteness we study the
$\nu = 12/17$ state, although all four states $\nu = 12/7,$ 12/17, 12/31,
12/41 have the same neutral sector.
Each of these states has an infinite number
of $|K({\bf m})| = 2$ operators.  For the $\nu = 12/17$ state,
$K({\bf m}) = 2$ operators
are relevant on compact regions and $K({\bf m}) = -2$ operators on noncompact
regions of the plane.  The maximally relevant points and contours are plotted in
Fig.~\ref{fig4} as functions of boost parameters. The points and the
intersections of the contours mark the position of fixed points.
The points marked $A$, $B$, $C$ are examples of the three different types of
fixed points.
Plotting the marginal contours of the $|K({\bf m})| = 2$ operators
gives Fig.~\ref{fig5}a-c.  Fig.~\ref{fig5}a was
obtained by choosing a basis to bring a point ($A$) of
sixfold symmetry to the origin.  There are also
points of fourfold symmetry ($B$) as at the origin of Fig.~\ref{fig4},
Fig.~\ref{fig5}b, and Fig.~\ref{fig6},
and points of twofold symmetry ($C$) as in Fig.~\ref{fig5}c.
There is no
{\it a priori} reason to favor one type over the others.  In the same way,
Fig.~\ref{fig2} could have been drawn using a different basis to bring
point $B$ at the origin.  The
third type of fixed
point has one operator maximally relevant and four marginal operators:
these points are visible in Fig.~\ref{fig5}a-c
as the crossings of four marginal
lines at the center of a marginal circle.  These ``double marginal''
fixed points resemble the fixed points of the Fibonacci $\nu = 5/3$ state
except that there are four rather than two marginal operators.
Fig.~\ref{fig6} shows a curious property of these four-condensate
edges: the most relevant contours plotted as functions of
``velocity'' coordinates rather
than ``momenta'' $p_i$ in (\ref{sec:boostdef}) turn out to be straight lines.
Marginal contours are not
straight lines, even those which are straight lines in Fig.~\ref{fig5}b.
Mathematically the most relevant contours are straight
because the square-root terms cancel in the
equation $\Delta({\bf m}) = 1$ which determines the contour, leaving only
linear terms.

The complicated patterns in Fig.~\ref{fig5}a-c have physical
consequences.  The sixfold symmetric
points like $A$ have three maximally relevant operators and an $SU(3)$ symmetry
identical to that of the $\nu = 3/5$ fixed point previously studied.
The fourfold
symmetric points like $B$
have two independent $|K({\bf m})| = 2$ operators and an
$SU(2) \times SU(2)$ symmetry which is similar to the
$SU(2)$ symmetry of the $\nu = 2/3$ fixed point.  The double marginal
points like $C$ are
shown in Section V to give a different tunneling exponent than
the roughly similar
$\nu = 5/3$ fixed point.  These different phases within the charge-unmixed
plane are important even if quantum Hall systems necessarily have
quantized conductance, as has been suggested.~\cite{haldane3}
Points $A$ and $B$ are stable and solvable but are shown in Section V
to have different measurable properties, so a single
FQH edge with impurities
can have several physically different stable phases.

A complete understanding of these dim $K = 4$ states would require
studying the three- or four-dimensional plots of which Fig.~\ref{fig5}a-c are
sections.  One difference between the dim $K = 4$ states and the
states studied up to this point is that there are small regions
of the charge-unmixed plane on which only one operator is relevant, making
it less certain that points not on the plane but near one of these regions
would flow toward the plane as required for robust quantization.  The
dashed line between $A$ and $B$ in Fig.~\ref{fig5}a passes
through one such region.
Some experimental properties of the dim $K = 4$ states are
discussed in Section V.

\section{Symmetries of the edge}

This section discusses the effects of impurity scattering on
the symmetries in the $\chi$LL theories of various edges.
The restoration of symmetry by impurity scattering
will be shown to explain the patterns in the phase diagrams found in
Section III.
The $\chi$LL theory of a quantum Hall edge contains two matrices
$K$ and $V$ and a charge
vector ${\bf t}$ as described
in Section II.  The integer matrix $K$ may admit discrete symmetries,
which are described by integer matrices $M$ invertible over the integers
with
\begin{equation}
M^{\rm T} K M = K,\quad M^{\rm T} {\bf t} = {\bf t}.
\label{sec:symdef}
\end{equation}
Most velocity matrices $V$ do not have such symmetries.  Thus a
symmetry possessed by $K$ is in general broken by the $V$ terms in the
$\chi$LL action.

One result of KFP is that impurity
scattering can drive the velocity matrix to a fixed point where all the
symmetries of $K$ are symmetries of the full theory.  In this section
we show that, for the edges with infinitely many fixed points found
in section III, impurity scattering sometimes restores some but not all of the
symmetry of the $K$ matrix.  Because of this broken symmetry,
the different fixed points are like spin-up
and spin-down fixed points for an Ising ferromagnet below the transition
temperature: the Ising fixed points are carried into each other by
spin rotation, which is
a symmetry of the starting Hamiltonian but not of
the fixed points.  The infinitely
many impurity fixed points are carried into each other by symmetries of
$K$ which are not symmetries of $V$ at the fixed points.  The
broken-symmetry structure can be very rich, as in the case of the
$\nu = 12/17$ state, which has three different types of fixed points,
each breaking different symmetries of $K$.

The matrix $M$ in (\ref{sec:symdef}) gives a transformation
$\phi_i = M_{ij} {\tilde \phi_j}$
of the bosonic fields $\phi_i$ under which the action is
form-invariant.  The discrete symmetry transformation $M$ can reflect an
underlying continuous symmetry, as in the theory of the $\nu = 2/3, 3/5,
4/7, \ldots$ states, where the discrete symmetries of the $K$ matrix
reflect an $SU(n)$ symmetry of the field theory,
$n = {\rm dim}\ K$.~\cite{read}
It is easily seen that the symmetries $M$ of a given
$K$ matrix form a group with matrix multiplication as the group product.
The key difference between edges with a single impurity fixed
point and edges with infinitely many fixed points is that the
the former have finite symmetry groups, while the latter have infinite
symmetry groups.  As examples of the two types, we find the symmetries
of the $\nu = 3/5$ (finite) and $\nu = 5/3$ (infinite) edges.  The
results presented for $\nu = 5/3$ also apply to the other Fibonacci-type
edges: $\nu = 5/7,\ \nu = 5/13,\ \nu = 5/17.$
Section V
shows that the two different types of fixed points in the $\nu = 5/3$ edge have
different experimentally observable properties.  The $\nu = 12/17$
edge (likewise $\nu = 12/7,\ \nu = 12/31,\ \nu = 12/41$) is shown to have
three different types of fixed points related by a complicated symmetry
group.

The $\nu = 3/5$ state in the hierarchy basis has
\begin{equation}
K = \left(\matrix{1&1&0\cr 1&-2&1\cr 0&1&-2}\right),\quad {\bf t} = (1,0,0).
\end{equation}
One way to find the symmetries of $K$ is to start with transformations $W$
bringing $K$ to diagonal form and preserving ${\bf t}$, as were used in
Section III to obtain phase
diagrams.  Let $D$ be the matrix with diagonal elements $(1,-1,1)$.
If $W^{\rm T}KW$ is diagonal, then $M = WDW^{-1}$ is a symmetry of
$K$ with the property that $M^2 = I$, the identity.
The effect of $M$ is to use $W$
to go to independent fields ${\tilde \phi}_i$, change the sign of one field,
and then return to the original fields.
The problem is that $M$ is only integral for some choices of $W$.  One hopes
that by choosing different matrices $W_i$, one can find enough
integral $M_i$ to
generate the entire group of symmetries.  The $M_i$ are improper
since det $M_i = -1$;
the proper symmetry group contains only products of even numbers of $M_i$. 

For $\nu = 3/5$ two generators found using this
trick are
\begin{equation}
x = \left(\matrix{1&0&0\cr 0&1&0\cr 0&1&-1}\right),
\quad y = \left(\matrix{1&0&0\cr 1&-1&1 \cr 0&0&1}\right).
\end{equation}
The element $xy$ is a proper symmetry which generates a $120^\circ$
rotation of Fig.~\ref{fig1}, and as expected $(xy)^3 = I$.  The symmetry group has six elements: three proper
elements $\{I, xy, (xy)^2 = y^{-1} x^{-1}\}$ and three improper elements
$\{x,y,xyx\}$.  It is easy to check that these six elements are the full
symmetry group $G$.  The velocity matrix at the fixed point also has all
of these symmetries.  (For the sake of exactness, recall that the origin
of Fig.~\ref{fig1}
represents the set of all velocity matrices with certain values
of the boost parameters, as described in Section II.  There is an additional
RG flow of the other parameters in $V$ which makes the two neutral modes
have the same velocity.  Without this additional flow, only the boost
part of the velocity matrix would have the symmetry.)

One simple consequence of the symmetry at the $\nu = 3/5$ fixed point is
that the $V$-dependent scaling dimension matrix $\Delta$, which determines the
scaling dimension of the operator $O_{\bf m} = \exp(i m_j \phi_j)$ according
to $\Delta({\bf m}) = m_i \Delta_{ij} m_j$, has the same symmetries as
$K^{-1}$: $x \Delta x^{\rm T} = y \Delta y^{\rm T} = \Delta$.  Note that
$\Delta$ transforms like $K^{-1}$ rather than $K$ so its symmetries are
transposed symmetries of $K$.
At the fixed point $\Delta$ is invariant under all symmetries of $K^{-1}$
for any
edge with all neutral modes moving opposite the charge mode, as now shown.
These edges have fixed points where $K^{-1}$ and $\Delta$ are
both diagonal in some integral basis with first basis vector
${\bf e}_1 = {\bf t}$.  $K^{-1}$ has all diagonal entries negative
except for the first, and $2 \Delta = |K^{-1}|$ has all diagonal entries
positive.
Any vector ${\bf m}$ with charge $q = {\bf t} K^{-1} {\bf m}$ can be written
as ${\bf m} = a {\bf t} + {\bf n}$,
where ${\bf n}$ has charge zero (${\bf t} K^{-1} {\bf n} = 0$)
and $a = {\bf t} K^{-1} {\bf m} / {\bf t} K^{-1} {\bf t} =
q \nu^{-1}$.  Now with $K({\bf x}) = {\bf x} K^{-1} {\bf x}$,
\begin{equation}
2 \Delta({\bf m}) = 2 \Delta(a {\bf t} + {\bf n}) = a^2 K({\bf t}) - K({\bf n})
= {q^2 \over \nu} - K({\bf n}).
\end{equation}
Let ${\bf m^\prime} = M {\bf m}$ be the image of ${\bf m}$ under a symmetry
of $K^{-1}$.
Then $2 \Delta({\bf m}^\prime) = q^2 / \nu - K({\bf n}^\prime)
= q^2 / \nu - K({\bf n}) = 2 \Delta({\bf m})$, since ${\bf n}^\prime =M{\bf
n}$.  Thus $\Delta$ has every
symmetry of $K^{-1}$ for any charge-unmixed fixed point in a state with
all neutral modes opposite the charge mode.  Broken-symmetry fixed points
therefore appear only in states with neutral modes in both directions.
The same argument gives that at any charge-unmixed
fixed point where $K$ and $\Delta$ are diagonal,
\begin{equation}
2 \Delta({\bf m}) \geq q^2 / \nu
\label{sec:inequal}
\end{equation}
where $q$ is the charge of ${\bf m}$.
This inequality appears in the discussion of quasiparticles in Section V.

The same technique can be used to find the symmetries of $K$ for
the $\nu = 5/3$ Fibonacci-type edge shown in Fig.~\ref{fig2}.
Two elements of the symmetry group
are found from changing the sign of $(1,-1,0)$, which corresponds to
reflecting $x \leftrightarrow -x$ in Fig.~\ref{fig2},
and from changing the sign
of $(0,1,3)$, which corresponds to reflecting the $x$-axis through the
point $B$.  The resulting matrices are
\begin{equation}
u = \left(\matrix{0&1&0\cr 1&0&0\cr 0&0&1}\right),
\quad v = \left(\matrix{1&0&0\cr 0&2&3 \cr 0&-1&-2}\right).
\end{equation}
The difference between this case and the previous one appears when
$u$ and $v$ are multiplied to obtain other group elements.
The element $w \equiv uv$ is a proper symmetry of infinite order:
$I,w,w^2,\ldots$ are all different matrices and all symmetries of
$K$.  Each application of $w$ corresponds to translating Fig.~\ref{fig2}
 horizontally.
The Fibonacci property ${\bf m}_{n + 1} = {\bf m}_n + {\bf m}_{n-1}$
mentioned earlier is a consequence of symmetry under $w$.
The powers of $w$ and its inverse give the entire proper symmetry group,
which is isomorphic to $Z^{+}$, the group of integers
under addition.  The full symmetry group is isomorphic to the semidirect
product of $Z^{+}$ and the binary group $\{1,-1\}$.

At each fixed point $\Delta$ has a much smaller symmetry group than
$K.$  The only symmetry of $\Delta$ at a fixed point other than $I$
is the unique reflection which changes the sign of the operator
maximally relevant at the fixed point.
For example, $u$ is a symmetry of point $A$
($u \Delta_A u^{\rm T} = \Delta_A$)
but $v$ is not.  It is apparent from Fig.~\ref{fig2} that
some symmetry of $K^{-1}$ is broken at each fixed point because neutral
operators ${\bf m}_i$ with the same minimum scaling dimensions
$K({\bf m}_i) = 2$ have different actual scaling dimensions
$\Delta({\bf m}_i)$.  The matrix $w = uv$ is a symmetry of no fixed point,
but its effect is to move the system from one fixed point to the next:
$w \Delta_i w^{\rm T} = \Delta_{i + 1}$, where $i$ labels fixed points
{\it of the same type}, i.e., $w$ never takes maximally
relevant points of $K({\bf m}) = -2$ operators to maximally relevant
points of $K({\bf m}) = 2$ operators, since $w$ preserves $K$.  Thus in
Fig.~\ref{fig2} there is no symmetry taking point $A$ to point $B$.  In Section
V it is shown that the two different types of fixed points have
different experimentally measurable properties.

By applying symmetries of $K$, the boost part of any velocity matrix can
be made to lie in the region bounded by the maximally relevant lines
of $(1,-1,0)$ and $(-1,2,3)$ in Fig.~\ref{fig2}.  This region is a
``fundamental period'' of the symmetries of $K$.  However, different
fixed points of the same type may correspond to experimentally
different phases, even though they are related by a discrete symmetry and will
have the same scaling dimensions, etc.  The reason is that an experimental
probe will couple nonuniversally to some combination of the original
fields $\phi_i$, which after applying a symmetry of $K$ will be some
different combination of the redefined fields $\phi^\prime_i$.  Experiments
will measure different prefactors for various quantities at
different fixed points of the same type.  Hence even if only points of
type $A$ are found to be stable for $\nu = 5/3$, for example, there would
still be multiple edge phases with true transitions at phase boundaries.
This is not true if there are {\it continuous} rather than discrete
symmetries of the $\chi$LL system relating fixed points of the same type,
since then all the fixed points are continuously connected.  Such a situation
occurs if the discrete symmetries of the bosonized $(K,V)$ theory
arise from continuous symmetries of the underlying fermionic Lagrangian.
We discuss this point further for the $\nu = 3/5$ state in Section VI.
Stable fixed points of different types always give different phases.

Multiple-condensate edges have quite complicated symmetry groups, and it
is an interesting mathematical exercise to classify these groups in terms
of familiar finitely generated groups.  The symmetry
group of $\nu = 3/5$ found above is $D_3$, the triangular dihedral group,
for example.  Principal hierarchy
states with all neutral
modes opposite the charge mode have finite symmetry groups, and principal
hierarchy states with
neutral modes in both directions have infinite symmetry
groups.  Non-principal hierarchy states often have no nontrivial symmetries.
Here we will be content to mention some
results on the four-condensate principal hierarchy
states discussed previously.
The four-condensate states $\nu = 12/7$, 12/17, 12/31, 12/41
have three distinct types of fixed points ($A,B,C$ in Fig.~\ref{fig5}a-c).
The phase diagram has sixfold symmetry about point $A$, fourfold symmetry
about point $B$, and twofold symmetry about point $C$.  It seems likely
that these point symmetries are sufficient to generate the full symmetry
group, which at point $A$ is broken to a six-element subgroup and similarly
for $B$ and $C$.  A fundamental
period of the symmetry group is drawn in Fig.~\ref{fig5}a.
A set of generating matrices for $\nu = 12/17$ in the hierarchy basis
is then
\begin{eqnarray}
m_1 &=& \left(\matrix{1&0&0&0\cr 0&1&0&0\cr 0&0&1&0\cr 0&0&1&-1}\right),\ 
m_2 = \left(\matrix{1&0&0&0\cr 1&-1&1&0\cr 0&0&1&0\cr 0&0&0&1}\right),
\nonumber \\
&\phantom{=}& \quad
m_3 = \left(\matrix{1&0&0&0\cr 0&1&0&0\cr 0&-1&-1&-1\cr 0&0&0&1}\right).
\end{eqnarray}
These $m_i$ were obtained with the sign-flip procedure used above:
for each $i$ det $m_i = -1$ and ${m_i}^2 = I$.
The symmetries of point $B$ are generated by $m_1$ and $m_2$, which commute,
and $m_3$ gives a rotation by $\pi$ around point $C$.
A sample element of order 3 is $m_1 m_3 m_2 m_3$, and an element of
infinite order is $m_3 m_2.$

\section{Implications for experiment}

The conductance and other experimental properties of a quantum Hall
state are affected by disorder according to the RG flows described in
the preceding sections.  One important feature of the three- and
four-condensate
principal hierarchy states is that they can have multiple phases within
the charge-unmixed subset of velocity matrices.  This is different
from the situation in two-condensate states and for any state with all
neutral modes moving in the same direction, where the quantization of
conductance occurs at a single point in boost-parameter space
and no phase transitions are predicted within the charge-unmixed
subset of velocity matrices.

In this section we first consider the $\nu = 5/3$ state and argue
that experimental setups are likely to be close to point $B$ in the
phase diagram, Fig.~\ref{fig2}.  The $\nu = 5/7$ state probably offers the best
chance for an experimentally accessible phase transition.  We calculate
electron and quasiparticle tunneling exponents for the different types
of fixed points found in the preceding sections and show that different
phases at the same filling fraction have different temperature dependences
of electron tunneling through a barrier.

The $\nu = 5/3$ state seen experimentally is likely to contain
both up spins and down spins: it consists of a $\nu = 1$ state of spin-up
electrons and a $\nu = 2/3$ state of spin-down electrons, or vice versa.
The fully polarized state has higher energy than the mixed-spin state
because some electrons lie in the second Landau level rather than the first,
costing energy proportional to $\hbar \omega_c$, $\omega_c$ the cyclotron
frequency.  This dominates the
savings in the Zeeman and Coulomb energies from polarizing the spins,
at least in GaAs, where the effective $g$-factor and Zeeman energy
are small.  The fully polarized state might appear in other materials
with larger $g$, or in tilted-field configurations which allow the Zeeman
energy to be increased with $\omega_c$ constant.

In the mixed-spin $\nu = 5/3$ state, scattering between up and down spins
is expected to be very weak unless magnetic impurities are added.  Thus the
spin-up and spin-down components are largely independent.  Independent
$\nu = 1$ and $\nu = 2/3$ liquids are described by point $B$
in Fig.~\ref{fig2}
because the velocity matrix which has no interactions between the two
liquids gives the scaling dimension matrix
\begin{equation}
2 \Delta = \left(\matrix{2 \Delta_1 & \matrix{0&0} \cr \matrix{0\cr0} &
2\Delta_{2/3}}\right)
= \left(\matrix{1&0&0\cr 0&2/3&0 \cr 0&0&2}\right),\quad {\bf t} = (1,1,0)
\end{equation}
which is brought by a change of basis to point $B$.  It is shown below that
point $B$ has the same low-temperature tunneling conductance exponent
$G \sim T^0$ as a combination of a $\nu = 1$ state ($G \sim T^0$) and
$\nu = 2/3$ state ($G \sim T^2$) would have.  The fixed point $A$ 
is not easily interpreted as a sum of two independent edges.  At $A$
the operator $(1,-1,0)$ which hops charge between the two right-moving
modes is maximally relevant, suggesting that in this phase the
$\nu = 1/3$ left-moving mode pairs with a bound, $SU(2)$ symmetric
combination of right-moving modes rather than with just one right-moving
mode as at point $B$.

The $\nu = 5/7$ ground state is spin-polarized and its two edge
fixed points may be more easily found experimentally than
those of the $\nu = 5/3$ state.
The $\nu = 5/7$ state is equivalent in $K$-matrix terms to a $\nu = 2/7$
gas of holes in a $\nu = 1$ state:
$K_h = M^{\rm T} K^\prime M, M^{\rm T} {\bf t^\prime} = {\bf t}$, with
${\bf t^\prime} = (1,1,0)$, ${\bf t} = (1,0,0)$,
\begin{eqnarray}
K_h &=& \left(\matrix{1&1&0\cr1&-2&1\cr0&1&2}\right), \nonumber \\
K^\prime &=& \left(\matrix{K_1 & \matrix{0&0} \cr
\matrix{0 \cr 0} & -K_{2/7}}\right) = \left(\matrix{1&0&0\cr
0&-3&-1\cr0&-1&2}\right), \nonumber \\
M &=& \left(\matrix{1&1&0\cr0&-1&0\cr0&0&1}\right).
\end{eqnarray}
However, the $V$ matrix $V_{1 - 2/7}$ with no interactions between the
$\nu = 2/7$ holes and $\nu = 1$ electrons gives a conductance (in
units of $e^2 / h$)
$\sigma = 9/7 = 1 + 2/7$ rather than $\sigma = 5/7 = 1 - 2/7$.  This happens
for exactly
the same reason that a $\nu = 2/3$ state with velocity matrix describing
$\nu = 1/3$ holes not interacting with $\nu = 1$ electrons gives a conductance
$\sigma = 4/3$: the quantized value of conductance is only obtained if
the edge equilibrates and all charged eigenmodes move in the same direction.

It is not difficult to find the point represented by
$V_{1 - 2/7}$ in the $\nu = 5/7$ version of Fig.~\ref{fig2}
(which looks similar
but with some stretching along the $y$-axis): it lies on the $y$-axis
with boost coordinates $(0,\sqrt{2/5})$.  This is not a fixed point
in the presence of disorder, and we expect the system to flow to a fixed
point of type $A$ or type $B$.  Unlike in the $\nu = 5/3$ case, where
type $B$ was easily interpreted as a $\nu = 1$ state plus a $\nu = 2/3$
state with no interactions between the two, for $\nu = 5/7$ we have
no simple interpretation of either phase as two independent
subedges.
The $K$ matrix $K_h$
is inequivalent
to a combination of $\nu = 2/3$ and $\nu = 1/21$ because det $K_h \neq
({\rm det}\ K_{2/3}) ({\rm det}\ K_{1/21})$, so no invertible integral
basis change can relate the two.  Below we show that the
$A$ and $B$ phases
can be distinguished experimentally, so that measurements of a $\nu = 5/7$
sample edge would allow its phase to be determined.  Then changes in
the $V$ matrix (from changes in the gate voltages, e.g.) might drive a new
type of impurity phase transition.

Before calculating tunneling properties for the various fixed points,
we would like to suggest briefly an experimental approach to
edge impurity scattering based on the existence of spin-polarized
and spin-singlet states at $\nu = 2/3$.  At $\nu = 2/3$ there is
an unpolarized spin-singlet state with the same
$K$ matrix and charge vector as the well-known spin-polarized state.
The polarized state is naturally interpreted as the particle-hole
conjugate of the
Laughlin $\nu = 1/3$ state~\cite{laughlin}, while the unpolarized state
is {\it not} the double-layer state consisting of a spin-up $\nu = 1/3$
state and a spin-down $\nu = 1/3$ state, which has an inequivalent $K$
matrix.  The unpolarized state can be studied in tilted-spin experiments
such as those of Eisenstein {\it et al.}~\cite{eisenstein} and appears
because of the relatively low Zeeman energy in GaAs as suggested by
Halperin.~\cite{halperin3}  The KFP treatment should be just as valid for the
unpolarized edge as for the polarized edge because they have the same
$K$ matrix.  The unpolarized edge has an exact $SU(2)$ symmetry if the Zeeman
energy is ignored, however, and this symmetry has physical consequences.

Numerical results on the unpolarized edge show that at low energy there
are two branches of excitations, one spin-singlet charge branch and one
spin branch described by the $SU(2)$ Kac-Moody algebra.~\cite{moore}
This is the structure found at the KFP fixed point and different
from the numerical results on the clean polarized edge, which indicate two
spatially separated subedges with no special symmetry.~\cite{johnson,wenrev1}
It seems logical that the physical requirement of $SU(2)$ spin
symmetry of the unpolarized edge forces the system to the KFP fixed point
even in the absence of disorder, assuming the ``hidden'' $SU(2)$ symmetry
is only found at the fixed point.  The $SU(2)$ structure of the unpolarized
edge is found in a small system (hence without RG flows)
for both Coulomb and short-range interactions.  The separation
of the $\nu = 2/3$ edge into charge modes and neutral modes can thus be
caused by (i) an exactly charge-unmixed velocity matrix, (ii) an
unbroken $SU(2)$ symmetry, or (iii) random impurities.  The
possibility that impurities affect the
polarized edge but not the unpolarized edge suggests that measurements
of the edge equilibration length and tunneling conductance across the
topological phase transition~\cite{mcdonald} between the two may be
illuminating.

In FQH states the tunneling conductance
through a point constriction in a Hall bar
decreases with decreasing temperature.  In
the integer effect this conductance is
temperature-independent.
The physical electron operator is a superposition of all charge-$e$
fermionic operators, and the low-temperature conductance is determined
by the scaling
dimension $\Delta_e$ of the most relevant such operator according
to~\cite{moon,kane2}
\begin{equation}
G(T) \approx t^2 T^{2(2 \Delta_e - 1)}
\end{equation}
where $t$ is the amplitude for the dominant tunneling process.
Different fixed points in the same FQH state can have different $\Delta_e$
and different tunneling exponents.  These exponents can be calculated for
the marginal-type fixed points even though the electron dynamics at these
points is unclear.  All fixed points of the same type have the same scaling
exponents but are expected to have measurably different prefactors
as described in Section IV.

	Charge-$e$ operators ${\bf m}$ have $t_i K^{-1}_{ij} m_j = 1$ and
scaling dimension $\Delta_e = m_i \Delta_{ij} m_j$ where $\Delta$ is the
same symmetric matrix calculated in Section III.  Since $\Delta$ is known
at each fixed point, it is simple to search for the most relevant
charge-$e$ operator.  The $SU(n)$ fixed points found by Kane and Fisher
for the $\nu = n/(2 n - 1)$ states have $2 \Delta_e = 3 - 2 n^{-1}$
and tunneling exponent
\begin{equation}
G(T) \approx t^2 T^\alpha, \quad\alpha = 4 - 4 n^{-1}.
\end{equation} 
In Table \ref{table2}
we list the low-temperature conductance behavior for each of
the fixed points found in Section III.  Note that corresponding fixed
points in states with the same neutral sector, such as $\nu = 5/3$
and $\nu = 5/7$, can have different tunneling exponents because the charge
sectors of the two edge theories are different.  The Fibonacci-type
states have two possible values of the low-temperature tunneling
conductance exponent, so that there is a real physical difference
between the $A$ and $B$ phases.

The level four states studied ($\nu = 12/7$, 12/17, 12/31, 12/41) have
three different tunneling exponents corresponding to the three different
types of fixed points.  For example, in the $\nu = 12/17$ state
the $SU(3)$ fixed points have $\Delta_e = 7/6$ and
$\alpha = 8/3$ as appear in the $SU(3)$ fixed point of the $\nu = 3/5$ state.
The $SU(2) \times SU(2)$ fixed point is the same as the $SU(2)$ fixed point
for $\nu = 2/3$ except that there are two charge-$e$ operators of minimal
scaling dimension rather than one.  The double marginal fixed point
has an operator with $\Delta_e = 11/12$ so $\alpha = 5/3$.  So the three
different fixed points have three different values of $\alpha$: 5/3 for
the double marginal points, 2 for the $SU(2) \times SU(2)$ points, and
8/3 for the $SU(3)$ fixed points.

Other tunneling experiments are sensitive to the most relevant
quasiparticle operator at a fixed point, rather than the most relevant
electron operator.  One experiment sensitive to the quasiparticle
scaling dimension is tunneling through a slight constriction rather than
through a deep constriction as described above.~\cite{kane}
We have calculated the scaling dimension of the most relevant quasiparticle
operators for the various fixed points.  No simple patterns are observed:
often two or more quasiparticle operators have nearly the same minimum
scaling dimension, and the charge of the most relevant quasiparticle
operator varies among different fixed points of the same edge.
As an example, in the $12/17$ edge the most relevant quasiparticles
at the different fixed points are:
$2 \Delta= 5/17 ,\ q=3 e / 17$ at the $SU(3)$ points,
$2 \Delta = 6/17,\ q=2e / 17$ at the $SU(2)$ points,
and $2 \Delta = 43/102,\ q = e/17$ at
the double-marginal points.  Typically the most relevant quasiparticles
have small charges, as expected from the inequality (\ref{sec:inequal}).

Time-domain experiments have so far not resolved the neutral modes in
nonchiral edge states,~\cite{ashoori} but in principle a perturbation
at one contact on a sample edge should excite propagating
charged and neutral modes observable at another contact.  Such
an experiment might reveal whether the neutral modes in the Fibonacci-type
states $\nu = 5/3, 5/7, 5/13, 5/17$ propagate or are localized.
The measurement of edge equilibration lengths might also give interesting
results:
measurements on the edge of the $\nu = 4/5$ edge, which has no
$|K({\bf m})| = 2$ operators and hence no
KFP-type instability, could show another type of equilibration mechanism
(such as inelastic scattering from phonons) with a different temperature
dependence.

\section{Summary}

We have developed a technique for studying impurity scattering
in a general FQH edge and used it to find phase diagrams and experimentally
measurable properties for a broad class of nonchiral edges.
We find that some FQH edges can have several different phases (fixed
points) in the presence of randomness.  These phases in general have higher
symmetry at low energies and long wavelengths than the original system.
Thus random edges demonstrate an interesting phenomenon of dynamical
restoration of symmetries at low energies and long length scales.
Different phases have different
experimentally observable properties.  It would be very interesting to find
these phases and study transitions between them experimentally.

The transitions between phases are interesting from the point of view
of Landau's symmetry breaking principle for continuous transitions:
A continuous phase transition (second-order in the Ehrenfest classification)
can only occur between two phases which differ in symmetry, and the symmetries
of one phase are a subset of the symmetries of the other phase.  This
principle appears to be satisfied by all the transitions between definitely
stable fixed points in the edges we study.  The principle is satisfied
even though the RG flows for
some transitions (such as the $\nu = 2/3$ transition~\cite{kfp}) are
similar to those in the Kosterlitz-Thouless transition, which is not
clearly interpreted in terms of a broken symmetry.  The symmetry breaking
principle also has some implications for a possible phase transition
in the $\nu = 5/3$ state, which has two types of possibly stable fixed points.

\begin{figure}
\epsfxsize=5.0truein
\centerline{\epsffile{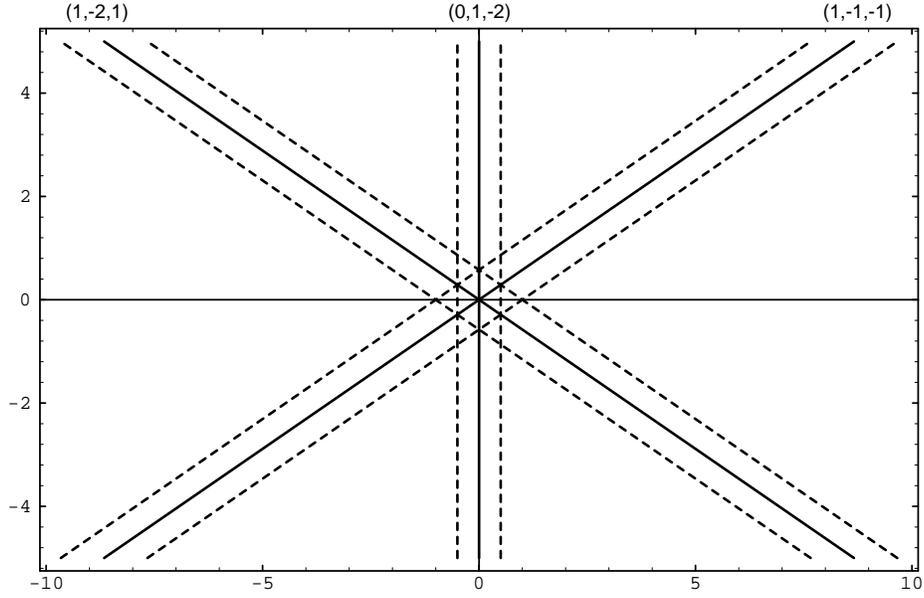}}
\caption{Plot of scaling dimension of the three $|K({\bf m})| = 2$
operators for
$\nu = 3/5$ edge as functions of boost parameters $(p_1,p_2)$.  The
charge-unmixed point is the origin.
Dashed lines indicate when operators become
marginal ($\Delta({\bf m}) = 3/2$) and
solid lines indicate when operators   become
maximally relevant ($\Delta({\bf m}) = 1$).}
\label{fig1}
\end{figure}

To summarize our results, two situations, with or without long-range
interactions, need to be discussed separately.
In the absence of long-range interactions, all the edge modes, in general,
carry some amount of charge and the edge is called charge-mixed. Several
different situations are illustrated by the following examples. 
\begin{itemize}
\item
The
$\nu=2/3$ edge has two phases. In one phase the edge is charge-mixed and the
two-terminal conductance $\sigma$ and the exponent $\alpha$ of electron 
tunneling between two edges $\sigma_{tun} \propto T^\alpha$ are not universal. 
In the other
phase the edge has an $SU(2)$ symmetry and
is charge-unmixed ({\it i.e.} only one propagating mode, the
charge mode, carries charge and other propagating modes are neutral).
In this case $(\sigma, \alpha)$ take universal values 
$(\frac23 \frac{e^2}{h}, 2)$. 
\item
The $\nu=3/5$ edge has three phases, described
by a fixed point (the point $(0,0)$ in Fig.~\ref{fig1}), fixed lines
(the solid lines 
outside the hexagon bounded by the dashed lines in Fig.~\ref{fig1}) and fixed
planes (the region outside the region bounded by the dashed lines). This is 
because a point on the fixed planes does not flow as the energy is lowered, 
while a point between two paralell dashed lines flows to the fixed line between 
them and a point inside the hexagon flows to the fixed point. The
fixed point has an
$SU(3)$ symmetry and is charge-unmixed. $(\sigma, \alpha)= 
(\frac35 \frac{e^2}{h}, \frac83)$ are universal. The fixed-line and fixed-plane
phases are charge-mixed and $(\sigma, \alpha)$ are not universal. But
in the fixed-line phase there is an $SU(2)$ symmetry and $(\sigma, \alpha)$
and other exponents all depend on a single parameter which parametrizes the
fixed line.  The fixed-plane phase has no particular symmetries.
We would like to point out that although the fixed-line phase
in Fig.~\ref{fig1}
contains six disconnected segments, this does not guarantee that there are six
disconnected fixed-line phases.  This is because the disconnected fixed lines
may be connected in a higher-dimensional space of Lagrangians
of which Fig.~\ref{fig1} is just a
two-dimensional cross section.  If different line segments are connected in
the enlarged space, it is possible to move continuously from one segment
to another without any transition.  For the $\nu = 3/5$ state the
higher-dimensional space results from applying the $SU(3)$ transformation
on the the full  Lagrangian. Note that the $SU(3)$ transformation does not
change the commutators between fermions (which can be seen in the fermionic 
form of the Lagrangian but is not evident in
the (Abelian) bosonized form), and 
hence leaves the Hilbert space unchanged.
Acting with the $SU(3)$ generators creates off-diagonal interaction
terms of the form $f(x) ({\bar \psi_1} \psi_2)
({\bar \psi_2} \psi_2)$, after we make the local $SU(3)$ transformation to
remove the random hopping term between different fermions.  Thus the precise 
form of the function $f(x)$ depends on the impurities which
generate the random hopping terms. 
If the off-diagonal terms have precisely the
variable coefficients $f(x)$, then the different fixed-line phases can be
continuously connected via inclusion of such off-diagonal terms. 
However, in real experiments it is impossible to control the
precise form of the variable coefficients $f(x)$, and the Lagrangians
for experimental samples
do not contain the above off-diagonal terms. Therefore for real samples all
different fixed-line phases are disconnected.
Similarly all fixed planes are disconnected for real samples.
However, since fixed-line phases (or fixed-plane phases) all have the same 
symmetry, the symmetry breaking principle prohibits
continuous phase transitions between two connected
fixed-line phases or two connected fixed-plane phases. But there are
still continuous phase transitions between
a fixed-line phase and a fixed-plane phase, and a fixed-line phase and a
fixed-point phase.  

We would like to stress that the sequence of the phase
transition: fixed-point phase $\to$  fixed-line phase $\to$ fixed-plane phase
represents a sequence of symmetry breaking: $SU(3) \to SU(2) \to SU(1)$.
This is consistent with the symmetry breaking principle discussed above.
It appears that the symmetry breaking principle
that governs the continuous transitions between clean 
phases in other condensed matter systems
also governs the continuous transitions between disordered phases of FQH edges.
All the continuous phase transitions between different disordered edge phases 
that we find in this paper are related to symmetry breaking.
\item
The $\nu=5/3$ edge also has three (types of) phases described
by fixed points (such as $B$ in Fig.~\ref{fig2}, but not $A$), fixed lines 
(the solid lines in Fig.~\ref{fig2})
and fixed planes (the region outside the region bounded by the dashed lines).  
Again the fixed-point phase is charge-unmixed and has universal 
$(\sigma, \alpha)$. The fixed-line and
fixed-plane phases are charge-mixed and have non-universal $(\sigma, \alpha)$.
However, here the fixed point contains two marginal operators. It is not clear
whether the fixed point is stable or not (depending on whether the two
marginal operators are marginally relevant or not).
It is not clear if different fixed lines and fixed planes are connected or not
in a higher-dimensional space.
There is a continuous phase transition between the fixed-line phase
(with $SU(2)$ symmetry) and fixed-plane phase (with no symmetry).
We note that both the fixed-point phase and the fixed-line phase have $SU(2)$
symmetry. According to the symmetry breaking principle for continuous
transition, either there is another phase separating the the fixed-point phase
and the fixed-line phase, or the fixed-point phase is unstable, or the
transition is first-order (discontinuous) and the first-order line
does not terminate in a
second-order point for any finite disorder strength.  The perturbative RG
in Appendix B is consistent with the last possibility.
\item
The $\nu=17/13$ edge ($K=(1,4,-4)$) again has three phases described by a
fixed point, a fixed line and a fixed plane. (Fig.~\ref{fig3})  However,
the phase
diagram is quite different from the above two. There can only be continuous
phase transitions between the following phases:
the fixed-point phase (with $SU(2)$ symmetry) 
$\Longleftrightarrow$ the fixed-plane phase (with no symmetry)
$\Longleftrightarrow$ the fixed-line phase (with $SU(2)$ symmetry).
\item
The $K=(l,2,-2,2)$ and $K=(l,-2,2,-2)$ edges are too complicated, and we will
only discuss them for the case of long-range interactions.
\end{itemize}

In the presence of long-range interactions the edge is always (nearly)
charge-unmixed and the two-terminal conductance always takes
the quantized value
$\sigma=\nu \frac{e^2}{h}$. We can restrict our discussion to the 
charge-unmixed 
subspace (the $(0,0)$ point in Fig.~\ref{fig1} and the $x$-axis in Fig.~\ref{fig2} and 3). 
Table II gives the low-temperature tunneling exponent for all the
charge-unmixed phases of principal hierarchy states.
The above examples with short-range
interactions can be easily modified to cover the case of long-range
interactions:
\begin{itemize}
\item
The $\nu=2/3$ edge has only one phase which is described by a fixed point.
$(\sigma, \alpha)$ take universal values $(\frac23 \frac{e^2}{h}, 2)$. 
\item
The $\nu=3/5$ edge has only one phase, described
by a fixed point (the point $(0,0)$ in Fig.~\ref{fig1}). 
The fixed-point phase is the same as the fixed-point phase for short-range
interactions: it has an $SU(3)$ symmetry and universal
$(\sigma, \alpha) = (\frac35 \frac{e^2}{h},\frac83)$.
\item
The $\nu=5/3$ edge has two (types of) phases described
by $A$-type and $B$-type fixed points in Fig.~\ref{fig2}.
The fixed-point phases have  universal 
$\alpha$ given by $\alpha=2/5$ for $A$-type and $\alpha=0$ for $B$-type. 
However both $A$-type and $B$-type
fixed points contain two marginal operators, 
and it is not clear whether the fixed points are stable.
\item
The $\nu=17/13$ edge ($K=(1,4,-4)$) has three phases described by two
fixed points ($A$ and $B$ in Fig.~\ref{fig3}) and a fixed
line (the $x$-axis outside the region
bounded by the dashed lines). All three phases are stable. The two fixed-point
phases have different universal values for the temperature exponent:
$\alpha=2$ for the $A$-type and $\alpha=18/17$ for the $B$-type
fixed-point phase.
$\alpha$ is not universal for the fixed-line phase.
The only continuous
phase transitions are between one of the two fixed-point phases
($SU(2)$ symmetry) and the fixed-line phase (no symmetry).
\item
The $K=(l,2,-2,2)$ and $K=(l,-2,2,-2)$ edges with $\nu=12/7, 12/17,...$ are 
very interesting. There are four different phases described by three types of
fixed points ($A$, $B$, $C$ in Fig.~\ref{fig4} and 5) and a fixed
line (the middle
segment of the dashed line connecting $A$ and $B$ in Fig.~\ref{fig5}a).
Certainly there are infinitely
many different disconnected $A$-, \hbox{$B$-}, $C$-type fixed points and
fixed lines in 
Fig.~\ref{fig5}a, and it is not clear if all fixed points (lines) of each type
are connected in a higher-dimensional
space. The $A$-type fixed point has an $SU(3)$ symmetry, the $B$-type fixed
point has an $SU(2)\times SU(2)$ symmetry, the fixed line and the $C$-type
fixed point have an $SU(2)$
symmetry. The exponent $\alpha$
has the universal values ($8/3$, $2$, $5/3$) for the
$A$-, $B$-, $C$-type fixed points respectively.  The $C$-type fixed
point has four
marginal operators and it is not clear whether it is a stable fixed point.
Among the three definitely stable phases the only
possible continuous transitions are:
$A$-type phases ($SU(3)$ symmetry) $\Longleftrightarrow$
fixed-line phase ($SU(2)$ symmetry) $\Longleftrightarrow$
$B$-type phases ($SU(2)\times SU(2)$ symmetry). These
transitions are consistent with the symmetry breaking principle for
continuous transitions.  An $A$-type $\Longleftrightarrow$ $B$-type transition
would violate the symmetry breaking principle and
is not found in the phase diagram.
\end{itemize}

This study just starts to reveal some general intrinsic structures of 
disordered phases of FQH edges and transitions between those phases.
It is amazing to see that different disordered phases are characterized by
symmetries and that phase transitions are characterized by
broken symmetries. Certainly there are many open problems and
much needs to
be done in order to have a complete theory of disordered edges. The
possibilities of transitions between different disordered phases on the
edge of a single bulk quantum Hall liquid also open up new directions for
experimental explorations.

J. E. M. wishes to thank W. R. Mann and C. Mudry for
helpful conversations.  
X.-G. W. is supported by NSF
Grant No. DMR--97--14198 and NSF-MRSEC Grant No. DMR--94--00334.
J. E. M. acknowledges a fellowship from the Fannie and John Hertz Foundation.

\appendix

\section{Boosts and Rotations}
\label{boostapp}
The decomposition of an element of $SO(m,n)$ into a product of a boost and
a rotation follows in a neighborhood of the origin simply by writing the
product $M = BR$ in terms of infinitesimal generators of the Lie group.
If $b_i$ are the boost generators and $r_i$ the rotation generators, then
a boost (similarly, rotation) close to the identity element contains only
boost (rotation) generators:
\begin{equation}
BR = (1 + \epsilon_i b^i) (1 + \delta_j r^j) \approx 1 + \epsilon_i b^i
+ \delta_j r^j,
\end{equation}
where $\epsilon_i$ and $\delta_j$ are arbitrary infinitesimal parameters.
There are exactly enough free parameters to cover a neighborhood of the
identity in $SO(m,n)$.  Thus if the decomposition does not hold on the
entire group, there must be some boundary in $SO(m,n)$ where it ceases
to hold.  In the next paragraph we outline a global proof of the decomposition.
The details are given for $SO(m,1)$, which is the only case used in the
body of this paper.

The boost part of a given matrix $M$ can be constructed if every
symmetric positive definite element of $SO(m,n)$
has a square root within the group which is also symmetric.
The square root is simple for
$m=1$ or $n=1$, where every symmetric positive definite matrix is of the
form introduced in Section II and associated with a unique velocity
vector ${\bf v} = {\bf p} / \gamma$.
Then the square root is the boost with velocity
${\bf v^\prime} = {\bf v} (1 - \sqrt{1 - v^2}) / v^2$,
which is chosen so that the special-relativistic velocity
addition formula holds:
${\bf v} = 2 {\bf v^\prime} / (1 + {v^\prime}^2)$.
For the general $SO(m,n)$ case, a square
root can be defined by the Inverse Function Theorem within a neighborhood
of the identity and analytically continued.
Such a square root exists globally if every boost matrix can be written as an
exponential of only boost generators, since then
\begin{equation}
B^\prime = \exp\left(\matrix{{\bf 0} & b \cr b^{\rm T} & {\bf 0}}\right),
\quad \sqrt{B^\prime} =
\exp\left(\matrix{{\bf 0}& {b \over 2} \cr {b^{\rm T} \over 2} &
{\bf 0}}\right).
\end{equation}

With a square root, the proof is simple.  Given an arbitrary element
$M \in SO(m,n)$, $M M^{\rm T}$ is symmetric and
positive definite, so let $B \equiv \sqrt{M M^{\rm T}}$.  It remains to show
that $R \equiv B^{-1} M$ is in $SO(m,n)$ and is orthogonal:
\begin{eqnarray}
R I_{m,n} R^{\rm T} &=& B^{-1} M I_{m,n} M^{\rm T} {B^{-1}}^{\rm T} 
\nonumber \\
&=& B^{-1} I_{m,n} {B^{-1}}^{\rm T} = I_{m,n},
\end{eqnarray}
\begin{eqnarray}
R^{\rm T} R &=& (\sqrt{M M^{\rm T}}^{\,-1} M)^{\rm T}
(\sqrt{M M^{\rm T}}^{\,-1} M) \nonumber \\
&=& M^{\rm T} (\sqrt{M M^{\rm T}}^{\,2})^{-1} M = I.
\end{eqnarray}

\section{Renormalization group treatment of the Fibonacci edge}
\label{rgapp}
The exactly solvable $SU(2)$ fixed point found by KFP is stable
under RG transformations: near the fixed point the
velocity matrix weakly couples the
modes which are decoupled at the fixed point, but this originally
marginal term in the action acquires a random coefficient from the
$SU(2)$ rotation and becomes irrelevant.  The fixed point is no longer
necessarily stable if there is an additional marginal disorder operator,
since this term has a random coefficient to begin with
and does not have its scaling dimension decreased by the $SU(2)$ rotation.
Therefore such a term remains marginal and must be treated.
In this section we obtain the first-order coupled RG equations
for the Fibonacci-type edge $\nu = 5/3$,
which has additional marginal operators present at each of its
fixed points.

These first-order equations suggest that
both types $A$ and $B$ of fixed points are
in fact stable and that which fixed point the system
flows to asymptotically depends on the initial velocity matrix and the
disorder strengths.  The picture from the first-order equations is
incomplete at the fixed point because the higher-order effects of one
disordered operator upon another are ignored, although
these effects may well determine the properties of the fixed point.  The main
conclusion of the somewhat complicated discussion below is that the
fixed points
with marginal operators are probably stable under RG transformations
although the long-length-scale dynamics at the fixed point are unclear.
Note that even if only one type of fixed point turns out to be stable,
different fixed points of that type are distinguishable phases
as discussed in Section IV.  Fixed points of the same type have the
same stability properties because they have the same $\chi$LL theory
up to a redefinition of fields.

For the calculation on the $\nu = 5/3$ edge, we will assume that the
system is on the charge-unmixed line and use the scaling dimension of
one operator $\Delta_1$ as a coordinate on this line.  This is reasonable
since points close to the line are driven to the line under RG by the
$K({\bf m}) = -2$ operators.  It is possible that the strengths of all
$K({\bf m}) = -2$ operators may decrease sufficiently rapidly that the system
is left on one of the $K({\bf m}) = 2$ marginal lines away from the charge-mixed
point.  Then there is only one relevant
operator and the fixed point is solvable with an $SU(2)$ symmetry, exactly
as for the $\nu = 2/3$ fixed point studied by KFP.  Only on the charge-unmixed
line does a new type of fixed point appear.

\begin{figure}
\epsfxsize=5.5truein
\centerline{\epsffile{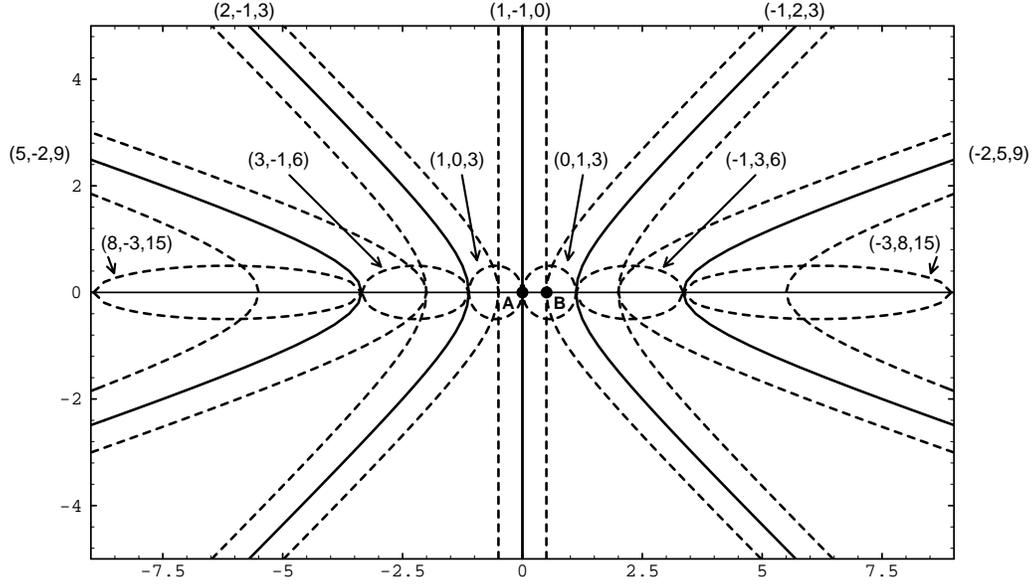}}
\caption{Plot of scaling dimension of the first 11
$|K({\bf m})| = 2$ operators for
$\nu = 5/3$ edge as functions of boost parameters $(p_n,p_c)$.
Dashed and solid lines are as in Fig. 1.
The charge-unmixed line is the $x$-axis.  At each point on the
$x$-axis where one operator is maximally relevant, two other operators
are marginal.  Points $A$ and $B$ are examples of the two different types
of fixed point.}
\label{fig2}
\end{figure}

\begin{figure}
\epsfxsize=5.0truein
\centerline{\epsffile{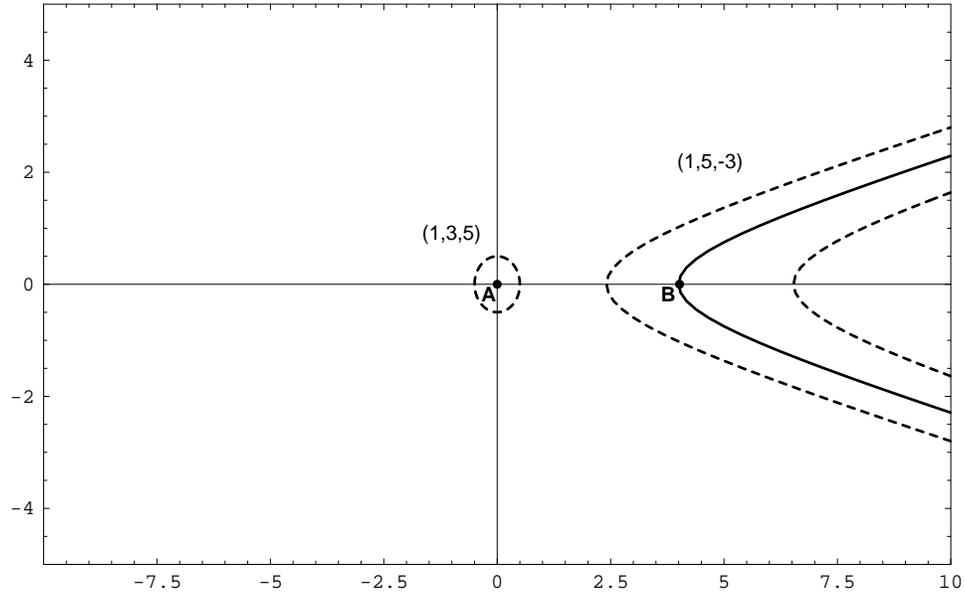}}
\caption{Plot of scaling dimension of the two $|K({\bf m})| = 2$ operators
for $\nu = 17/13$.  Axes are as in Fig. 2.  At most points on the
charge-unmixed line, there are no relevant disorder operators.  Points
$A$ and $B$ are the two charge-unmixed fixed points.}
\label{fig3}
\end{figure}

\begin{figure}
\vbox{\epsfxsize=5.0truein
\centerline{\epsffile{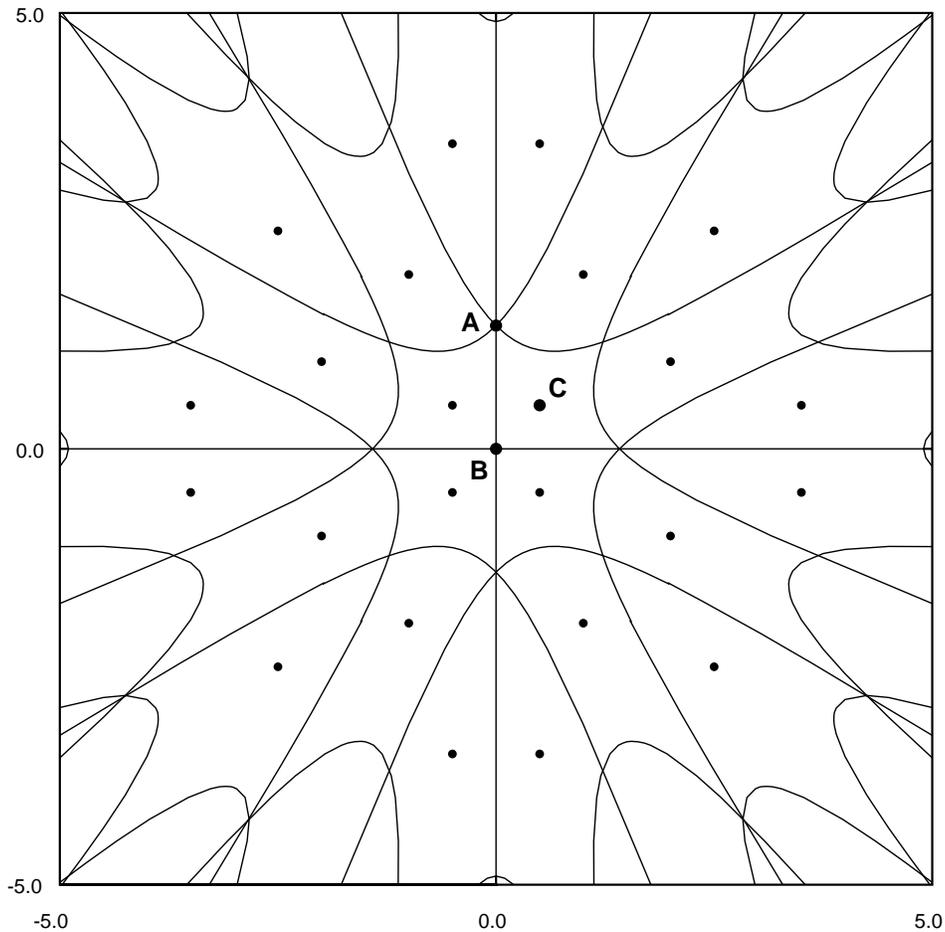}}
\caption{The most relevant contours of $|K({\bf m})| = 2$ operators
on the charge-unmixed plane of the $\nu = 12/17$ edge as functions of
boost parameters $(p_1,p_2)$.  Points $A$, $B$, $C$ are examples
of the three different types of fixed points: $A$ is an
$SU(3)$ point, $B$ an $SU(2) \times SU(2)$ point, and $C$ a
``double marginal'' point.
Dots are most relevant points of $K({\bf m}) = 2$ operators, lines are
most relevant lines of $K({\bf m}) = -2$ operators.}
\label{fig4}}
\end{figure}

\begin{figure}
\epsfxsize=5.0truein
\centerline{\epsffile{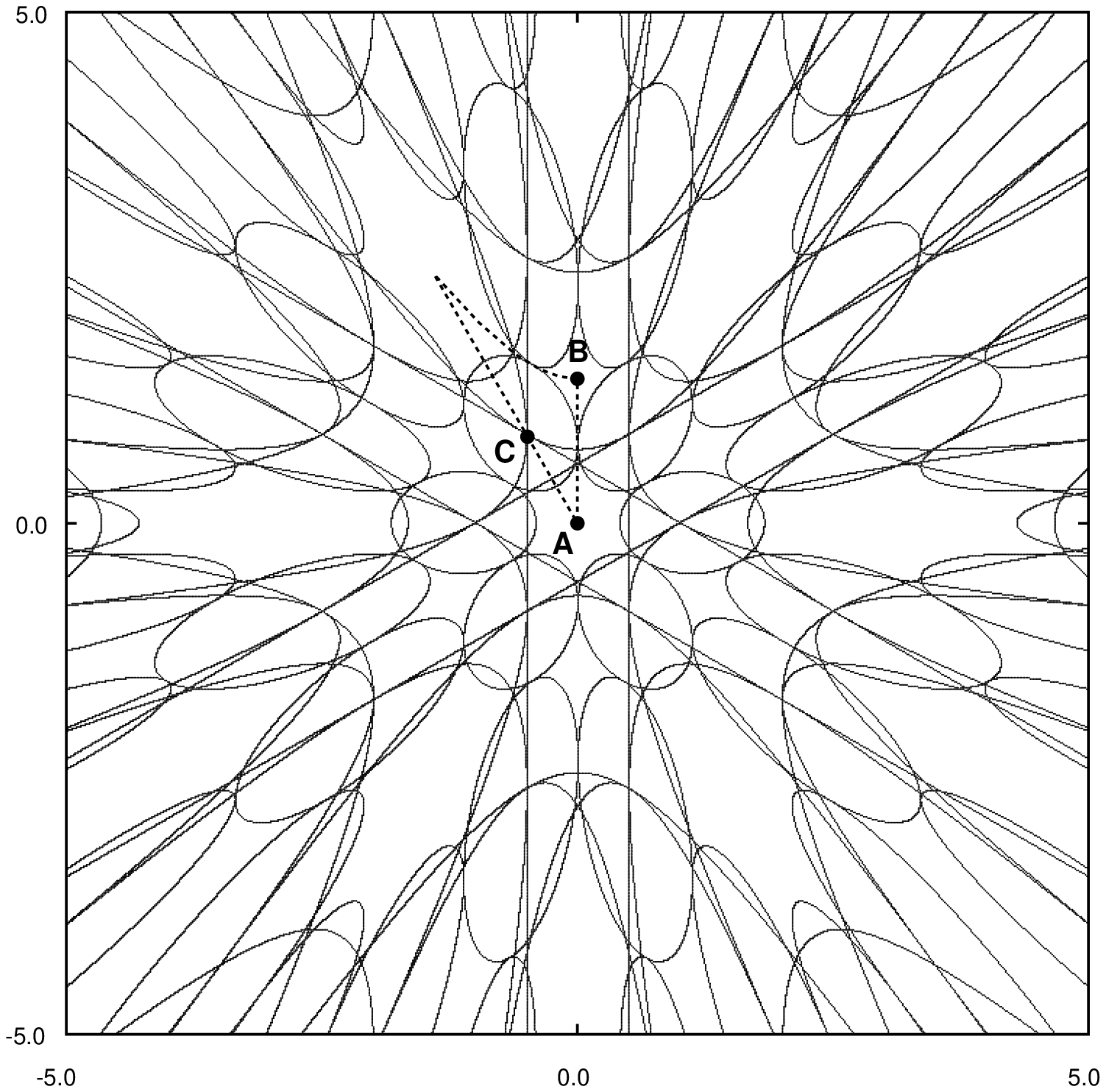}}
\epsfxsize=5.0truein
\centerline{\epsffile{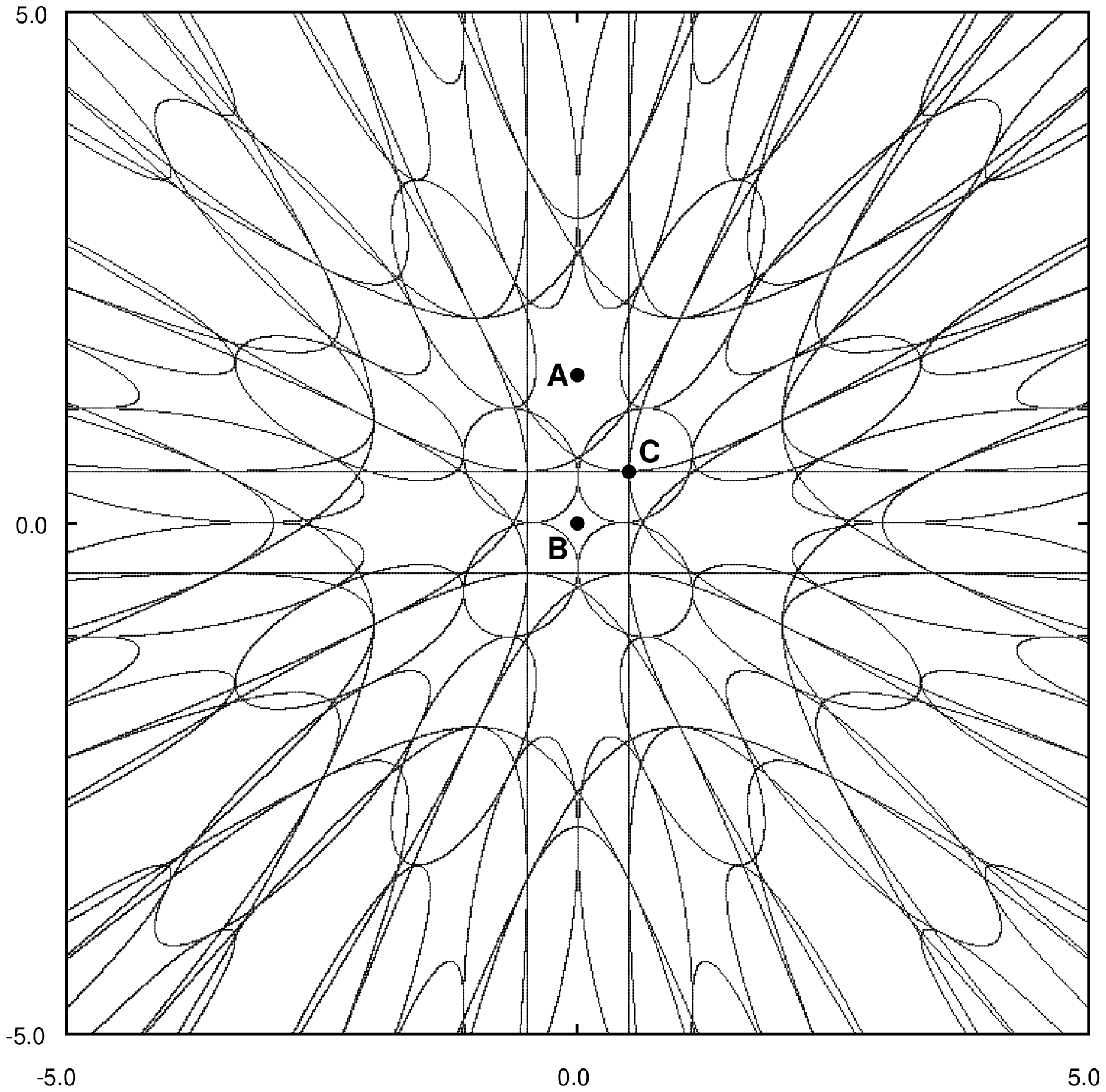}}
\epsfxsize=5.0truein
\centerline{\epsffile{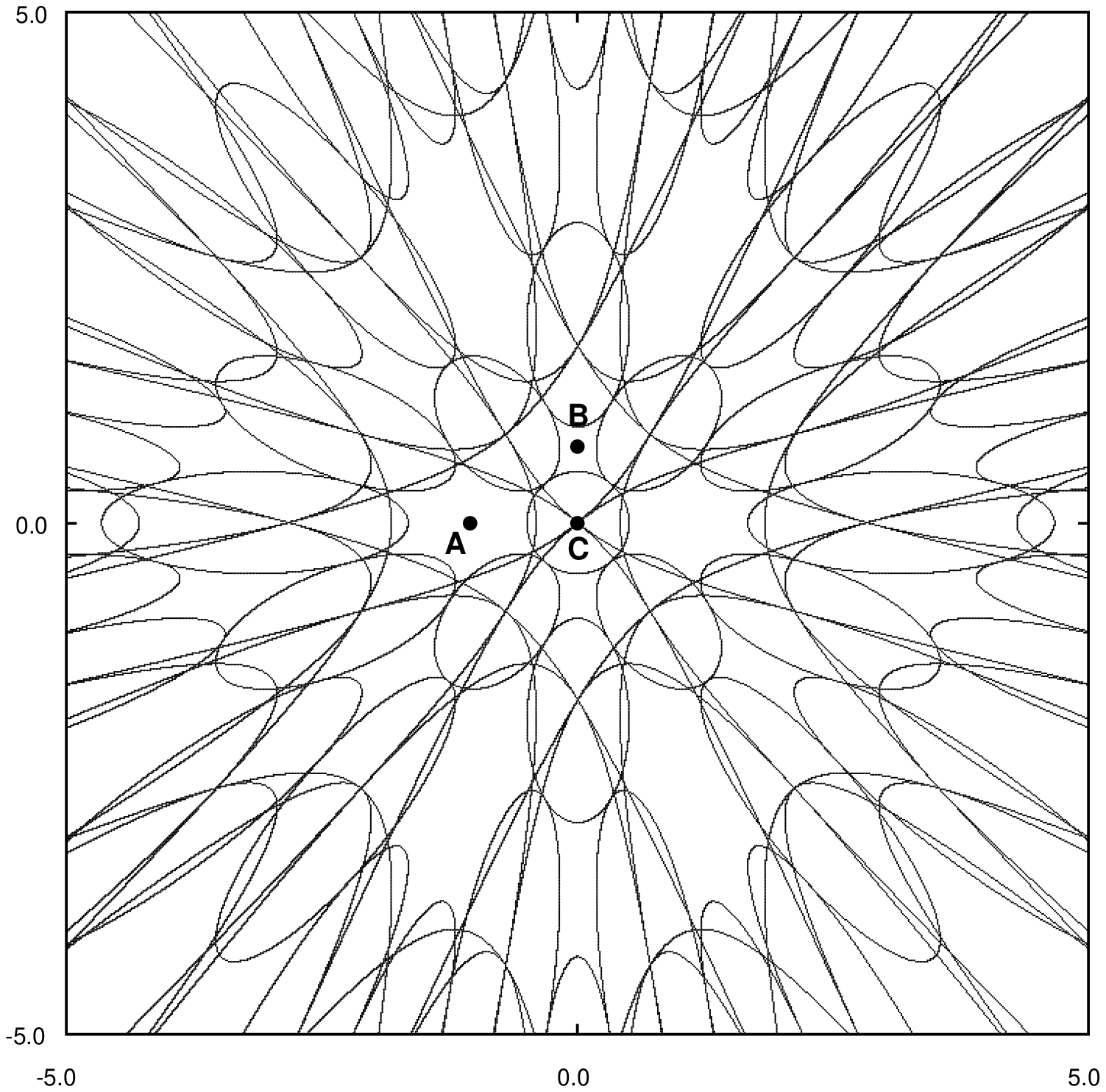}}
\caption{Plots (a-c) show the {\it marginal} contours
rather than the most relevant contours of $|K({\bf m})| = 2$
operators for the $\nu = 12/17$ edge.
The three plots were obtained using different bases: (a) has the $SU(3)$
point $A$ at the origin and (b) the $SU(2) \times SU(2)$ point $B$,
while (c) has the ``double marginal'' point $C$.  Note that the
three plots have the same topology.  Plot (b) is the same as Fig. 4
except that marginal rather than most relevant contours are shown.
}
\label{fig5}
\end{figure}

\begin{figure}
\epsfxsize=5.0truein
\centerline{\epsffile{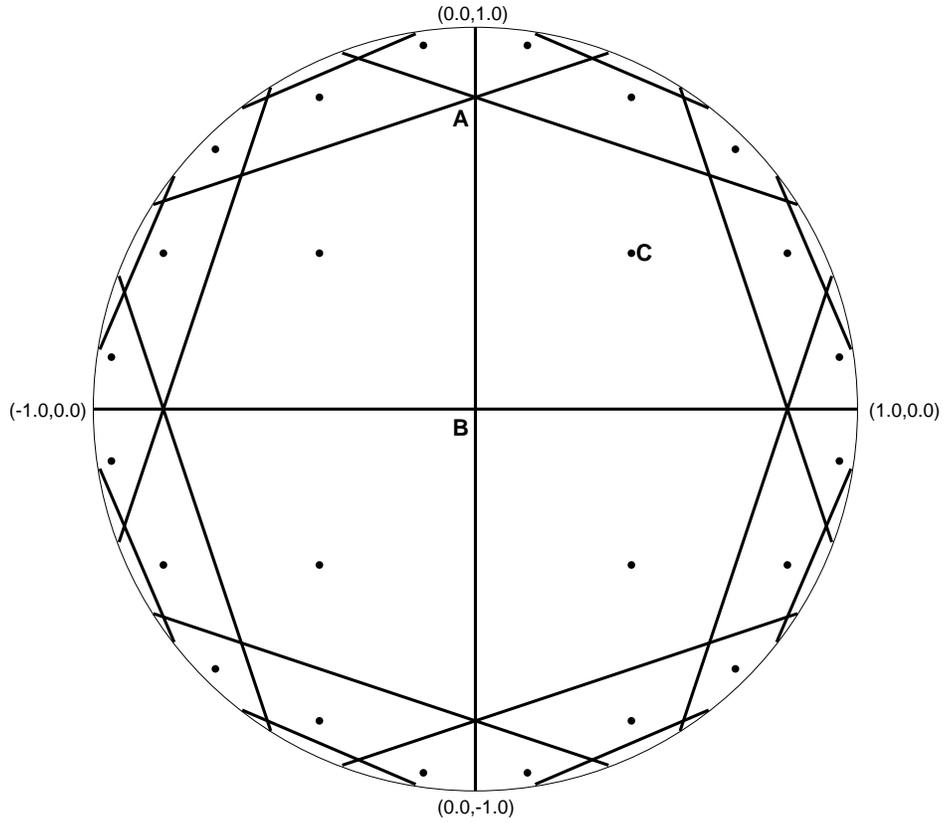}}
\caption{The most relevant contours of $|K({\bf m})| = 2$ operators on
the charge-unmixed plane of the $\nu = 12/17$ edge as functions of
{\it ``velocity''} coordinates $(v_1,v_2)$.  Plot is the same as Fig. 
4 except that contours are shown as functions of ``velocities'' rather
than ``momenta''.  Only the 42 most relevant operators at the origin are
shown because the full diagram becomes infinitely dense at the edge of the
circle.}
\label{fig6}
\end{figure}

At most three disorder
operators can be relevant at a point on the charge-mixed
line.  One operator's scaling dimension serves as an independent coordinate
and determines the other scaling dimensions: writing $\Delta_1$ for the scaling dimension of the operator whose
maximally relevant point will be studied, the scaling dimensions
of the two neighboring operators (Fig.~\ref{fig2}) are the two roots $\Delta_{\pm}$ of
\begin{equation}
\Delta_{\pm}^2 - 3 \Delta_{\pm} \Delta_1 + \Delta_1^2 + {5 \over 4} = 0.
\label{sec:sdim}
\end{equation}
The disorder strength of each operator has leading-order RG flow
$d D_i / d \ell = (3 - 2 \Delta_i) D_i$.  It remains to calculate how
the velocity matrix and scaling dimensions flow.

The decomposition of the velocity matrix in Section III into a boost part which
determines the scaling dimensions, plus a remaining ``rotational'' part, is
{\it not} compatible with the RG transformation.  The rotational part
affects the flow
of the boost part and vice versa.  However, the qualitative character of the
boost part is in some sense not affected by the rotational part as now
explained.  (Note that the $K$ matrix is invariant under the RG transformation
because it is purely topological and does not enter the Hamiltonian.)
Consider the perturbative RG flow equation for the scaling
dimension of the one $|K({\bf m})| = 2$ operator for the $\nu = 2/3$ edge:~\cite
{kfp}
\begin{equation}
{d\Delta \over d\ell}  =  -8\pi {\sqrt{v_+ v_-^{-3}}\over v_+ + v_-}
(\Delta^2 - 1)D.
\end{equation}
The eigenmode velocities $v_+, v_-$ are in the rotational part of the velocity
matrix and flow according to
\begin{equation}
{dv_\pm\over{d\ell}} = -4\pi {v_\pm^2 \over \sqrt{v_+ v_-^5}}
(\Delta \mp 1) D.
\end{equation}
The eigenmode velocities affect how fast $\Delta$ flows to 1,
but the basic idea that $\Delta$ flows smoothly to 1 is independent of
the precise values of $v_+$ and $v_-$ and of the details of their flow.
In order to make the coupled RG flows tractable we will
replace velocity-dependent prefactors by constants.

	Suppose that the system is in between the maximally relevant points
of two disorder operators with strengths $D_1, D_2$.  Then from
(\ref{sec:sdim}) the scaling dimensions
flow as
\begin{eqnarray}
{d \Delta_1 \over d \ell} &=& - c_1 (\Delta_1^2 - 1) D_1 \nonumber \\
& &+c_2 {3 \Delta_1 - 2 \Delta_2 \over 3 \Delta_2 - 2 \Delta_1} (\Delta_2^2 - 1)
D_2, \label{sec:sd1} \\
{d \Delta_2 \over d \ell} &=& - c_2 (\Delta_2^2 - 1) D_2 \nonumber \\
& &+ c_1 {3 \Delta_2 - 2 \Delta_1 \over 3 \Delta_1 - 2 \Delta_2} (\Delta_1^2 - 1)
D_1, \label{sec:sd2}
\end{eqnarray}
with $c_1$ and $c_2$ some positive constants.  These
two equations are not independent
and each together with (\ref{sec:sdim}) determines the other.  A complete
set of first-order equations consists of one scaling-dimension equation
plus the flow of the disorder strengths, with the other scaling dimension
found by (\ref{sec:sdim}).  The two disorder operators
compete to drive the system to one of the maximally relevant points.
The disorder strength flow implies that the fixed point which must exist
somewhere between the two maximally relevant points is unstable, as expected,
because on each side of this unstable point the disorder strength pushing
the system away is the more rapidly growing of the two disorder operators. 
What happens at one of the maximally relevant points is a little tricky,
because at such a point one disorder strength $D_1$ is growing
rapidly but
does not push the system in either direction since $\Delta^2 - 1 = 0$.

The main physical question to be settled is whether the disorder strength
of one of the marginal operators becomes infinite, remains finite,
or decreases to zero as the system moves to the fixed
point.  Now we show that in the
leading-order equations the marginal disorder strengths remain finite as the
system decays exponentially to the fixed point.  This does not necessarily
mean that the disorder strength of a marginal operator actually remains
finite, since exactly at the fixed point
this disorder strength is a constant to leading order but may increase or
decrease at higher order.  Linearizing
the flow equations about $\Delta_1 = 1, \Delta_2 = 3/2$, and keeping
$\Delta_2$ as independent variable and writing $\epsilon = 3/2 - \Delta_2 > 0$,
we have
\begin{eqnarray}
{d D_1 \over d \ell} &=& (1  - 4 \epsilon^2 / 5) D_1 \approx D_1, \\
{d D_2 \over d \ell} &=& 2 \epsilon D_2, \label{sec:d2eq} \\
{d \epsilon \over d \ell} &=& - c_2 (9 / 4 + 3 \epsilon - 1) D_2 + c_1
\epsilon D_1 \nonumber \\
&\approx& - 5 c_2 D_2 / 4 + c_1 \epsilon D_1. \label{sec:eps}
\end{eqnarray}
An asymptotic solution is found by taking $D_1 = D_1(0) \exp(\ell),
\epsilon = A \exp(-\ell).$  The right side of (\ref{sec:eps}) will balance
if $D_2 \rightarrow 4 c_1 A / 5 c_2$ as $\ell \rightarrow \infty$ since
the left side is much smaller in magnitude.  This limiting form is
consistent since with this form of $\epsilon$, (\ref{sec:d2eq})
yields a finite $D_2 = \exp(\int 2 \epsilon d\,\ell)$.  The linearized
analysis suggests that eventually the system decays exponentially to the
fixed point with $D_2$ asymptotically finite.
This prediction is confirmed by numerical integration of the
original differential equations.  Which fixed point the system reaches
as $\ell \rightarrow \infty$ depends the
initial scaling dimensions and disorder strengths.

A more complete RG treatment would give a deeper understanding of
the fixed points, but determining the RG equations to second order
in the disorder becomes quite complicated.  We discuss some features
here and hope to treat this problem more fully in another publication.
A preliminary step is to assume that the system is driven near
the fixed point by the first-order terms in the equation and then to carry
out an $SU(2)$ rotation to eliminate the random term with the
maximally relevant operator.  If the impurity operators are uncorrelated,
the marginal operators will still have random coefficients.  However, it is
physically likely that different impurity operators will be at least partially
correlated, possibly giving a uniform coefficient for the marginal operator,
which then becomes relevant.  In the case of uncorrelated impurities,
the possibility of carrying out this rotation shows that no terms involving
$D_1$ appear in the flow equation for $D_2$ exactly at the fixed point,
so there is still hope for a perturbative treatment.
Near but not at the fixed
point, the $SU(2)$ rotation of the dominant impurity operator may
affect the marginal operator even if the impurities are uncorrelated.
Even if the second-order terms in this
case have the proper sign to drive $D_2$ to zero, $dD_2/d\ell
\propto -D_2^2,$ the decrease of $D_2$ is only as $1 / \ell$ rather than
$\exp(-\ell)$, and at finite temperature the marginal operators should
have significant effects.
Another approach to understanding the marginal
fixed point is via an exact solution, which we have not been able to find.

\eject
\onecolumn

\onecolumn
\begin{table}
\mediumtext
\caption{Possible nonchiral edge types for dim $K \leq 4$.  For dim
$K = 5$ there are no principal hierarchy states and for dim $K > 5$
no states at all which are $T$-stable and have neutral modes in both
directions.}
\label{table1}
\bigskip
\bigskip
\vbox{ \offinterlineskip
\baselineskip=14pt
\halign{\vrule#\quad&\strut\hfil#\hfil&\quad
#\hfil&\quad#\hfil&\quad\hfil#\hfil
&\quad#\hfil&\quad\vrule#\cr
\noalign{\hrule}
\noalign{\vskip 2 pt}
\noalign{\hrule}
height4pt&\omit&\omit&\omit&\omit&\omit& \cr
&dim K&Mode directions&Example&Boost parameters&Boost parameters with&\cr
&&(charge always $\rightarrow$)&&&charge mode unmixed&\cr
height3pt&\omit&\omit&\omit&\omit&\omit&\cr
\noalign{\hrule}
height6pt&\omit&\omit&\omit&\omit&\omit& \cr
&2&\qquad\quad$1 \rightarrow \atop 1 \leftarrow$
&\ $\nu = 2/3$&1&\qquad\qquad 0&\cr
height10pt&\omit&\omit&\omit&\omit&\omit& \cr
&3&\qquad\quad$2 \rightarrow \atop 1 \leftarrow$
&\ $\nu = 5/3$&2&\qquad\qquad 1&\cr
height5pt&\omit&\omit&\omit&\omit&\omit& \cr
&&\qquad\quad$1 \rightarrow \atop 2 \leftarrow$
&\ $\nu = 3/5$&2&\qquad\qquad 0&\cr
height10pt&\omit&\omit&\omit&\omit&\omit& \cr
&4&\qquad\quad$3 \rightarrow \atop 1 \leftarrow$
&\ $\nu = 12/31$&3&\qquad\qquad 2&\cr
height5pt&\omit&\omit&\omit&\omit&\omit& \cr
&&\qquad\quad$2 \rightarrow \atop 2 \leftarrow$
&\ $\nu = 12/17$&4&\qquad\qquad 2&\cr
height5pt&\omit&\omit&\omit&\omit&\omit& \cr
&&\qquad\quad$1 \rightarrow \atop 3 \rightarrow$
&\ $\nu = 4/7$&3&\qquad\qquad 0&\cr
height6pt&\omit&\omit&\omit&\omit&\omit& \cr
\noalign{\hrule}
\noalign{\vskip 2 pt}
\noalign{\hrule}
}
}
\end{table}
\bigskip
\begin{table}
\caption{Low-temperature tunneling conductance behavior $G \sim T^\alpha$
for hierarchical daughter states of $\nu = 1$ (top) and $\nu = 1/3$.
Only ``charge-unmixed'' phases
(those with quantized conductance, or alternately those which can
occur with long-range interactions) are shown.
The different fixed points $A$ and $B$ for the Fibonacci-type states correspond
to the labeling in Fig. 2.
The phases $\nu = 12/7, 12/17, 12/31, 12/41$ have fixed lines $L$
and three types of fixed point with $SU(2) \times SU(2)$
symmetry (abbreviated $SU(2)$ in the table),
$SU(3)$ symmetry, or two independent marginal operators ($DM$).
The tunneling exponent on the fixed lines $L$ is nonuniversal.
Note that each exponent in the lower table is given by
$\alpha_{1/3} = 4 + \alpha_1$ where $\alpha_1$ is the exponent of the
state in the upper table at the same position in the hierarchy.
The pattern continues to lower filling fractions: daughter states
of $\nu = 1/5$ have filling fractions between $1/8 < \nu < 1/4$
and tunneling exponents $8 \leq \alpha \leq 12$, e.g.}
\label{table2}
$$\vbox{\tabskip = 0pt \offinterlineskip
\halign to 38em{\tabskip=0pt plus 1em#&
  #&#&\hfil#\hfil&#&\hfil#\hfil&#&\hfil#\hfil&#&#\hfil&#\tabskip=0pt\cr
&&&&&&\multispan3$\cdots \nu = n$\hfill&$G\sim T^0$\hfill\cr
&&&&&\strut&\vrule\cr
&&&&&\strut $\nu = 3, G \sim T^0$&\vrule&&&$G\sim T^{0}$&$DM$\hfill\cr
&&&&\multispan3\hrulefill\cr
&&&\strut $\nu = 2, G \sim T^0$&\vrule&&&$\nu = 12/7\quad$&&
$G\sim T^{1/3}$&$SU(2)$\hfill\cr
&&\multispan3\hrulefill&&\multispan3\hrulefill\cr
&&\vrule&&\vrule&\strut $\nu = 5/3,
G \sim {T^0 \atop
T^{2/5}}\,\,{{\rm B} \atop {\rm A}}$&\vrule
&&&$G\sim T^1$&$SU(3)$\hfill\cr
&\vphantom{.}&\vrule&&\vrule&&\vrule\cr
&&\vrule&&\multispan3\hrulefill\cr
&\strut&\vrule&&&&&&&$G\sim T^\alpha$&$L$\cr
&\strut $\nu = 1, G \sim T^0$ &\vrule\cr
\multispan3\hrulefill\vrule\cr
&&\vrule&&&\strut $\nu = 5/7, G \sim {T^{8/5} \atop T^2}\,\,{{\rm B} \atop
{\rm A}}$&&&&$G\sim T^{5/3}$&$DM$
\hfill\cr
&\vphantom{.}&\vrule\cr
&&\vrule&&\multispan3\hrulefill\cr
&&\vrule&\strut $\nu = 2/3, G \sim T^2$&\vrule&&\vrule&$\nu = 12/17$
&&$G\sim T^2$&$SU(2)$\hfill\cr
&&\multispan3\hrulefill&&\multispan3\hrulefill\cr
&&&&\vrule&\strut $\nu = 3/5, G\sim T^{8/3}$&&&&
$G \sim T^{8/3}$&$SU(3)$\hfill\cr
&&&&\multispan3\hrulefill\cr
&&&&&&\vrule&&&$G \sim T^\alpha$&$L$\cr
&&&&\strut&&\vrule\cr
&&&&&&\multispan3$\cdots \nu = \frac{n}{2 n - 1}$\hfill&$G \sim T^{4 - 4/n}$\hfill\cr
}}$$

$$\vbox{\tabskip = 0pt \offinterlineskip
\halign to 40em{\tabskip=0pt plus 1em#&
  #&#&\hfil#\hfil&#&\hfil#\hfil&#&\hfil#\hfil&#&#\hfil&#\tabskip=0pt\cr
&&&&&&\multispan3$\cdots \nu = {n \over 2 n + 1}$\hfill&$G\sim T^4$\hfill\cr
&&&&&\strut&\vrule\cr
&&&&&\strut $\nu = 3/7, G \sim T^4$&\vrule&&&$G\sim T^{4}$&$DM$\hfill\cr
&&&&\multispan3\hrulefill\cr
&&&\strut $\nu = 2/5, G \sim T^4$&\vrule&&&$\nu = 12/31$&&
$G\sim T^{13/3}$&$SU(2)$\hfill\cr
&&\multispan3\hrulefill&&\multispan3\hrulefill\cr
&&\vrule&&\vrule&\strut $\nu = 5/13, G \sim {T^4 \atop
T^{22/5}}\,\,{{\rm B} \atop {\rm A}}$&\vrule
&&&$G\sim T^5$&$SU(3)$\hfill\cr
&\vphantom{.}&\vrule&&\vrule&&\vrule\cr
&&\vrule&&\multispan3\hrulefill\cr
&\strut&\vrule&&&&&&&$G\sim T^\alpha$&$L$\cr
&\strut $\nu = 1/3, G \sim T^4$ &\vrule\cr
\multispan3\hrulefill\vrule\cr
&&\vrule&&&\strut $\nu = 5/17, G \sim {T^{28/5} \atop T^6}\,\,{{\rm B} \atop
{\rm A}}$&&&&$G\sim T^{17/3}$&$DM$
\hfill\cr
&\vphantom{.}&\vrule\cr
&&\vrule&&\multispan3\hrulefill\cr
&&\vrule&\strut $\nu = 2/7, G \sim T^6$&\vrule&&\vrule&$\nu = 12/41$
&&$G\sim T^6$&$SU(2)$\hfill\cr
&&\multispan3\hrulefill&&\multispan3\hrulefill\cr
&&&&\vrule&\strut $\nu = 3/11, G\sim T^{20/3}$&&&&
$G \sim T^{20/3}$&$SU(3)$\hfill\cr
&&&&\multispan3\hrulefill\cr
&&&&&&\vrule&&&$G \sim T^\alpha$&$L$\cr
&&&&\strut&&\vrule\cr
&&&&&&\multispan3$\cdots \nu = \frac{n}{4 n - 1}$\hfill&$G \sim T^{8 - 4/n}$\hfill\cr
}}$$
\end{table}

\begin{references}
\bibitem{klitzing}{K. von Klitzing, G. Dorda, and M. Pepper, Phys. Rev.
Lett. {\bf 45}, 494 (1980).}
\bibitem{laughlin}{R. B. Laughlin, Phys. Rev. Lett. {\bf 50}, 1395 (1983).}
\bibitem{halperin1}{B. I. Halperin, Phys. Rev. B {\bf 25}, 2185 (1982).}
\bibitem{milliken}{F. Milliken, C. Umbach and R. Webb, 
Solid State Comm., {\bf 97}, 309 (1995);
A.M. Chang, L.N. Pfeiffer and K.W. West, Phys. Rev. Lett, {\bf 77}, 2538 (1996).
}
\bibitem{ashoori}{R. C. Ashoori, H. Stormer, L. Pfeiffer, K. Baldwin and
K. West, Phys. Rev. B {\bf 45}, 3894 (1992).}
\bibitem{rev}{For a review, X.-G. Wen, Adv. in Phys. {\bf 44}, 405 (1995).}
\bibitem{zhang} {
S. M. Girvin and A. H. MacDonald, Phys. Rev. Lett. {\bf 58}, 1252 (1987);
S. C. Zhang, T. H. Hansson and S. Kivelson, 
Phys. Rev. Lett. {\bf 62}, 82 (1989); 
N. Read, Phys. Rev. Lett. {\bf 62}, 86 (1989);
Phys. Rev. Lett. {\bf 65}, 1502 (1990);
X.-G. Wen and A. Zee, Nucl. Phys. {\bf B15}, 135 (1990);
B. Blok and X.-G. Wen, Phys. Rev. {\bf B42}, 8133 (1990); 
Phys. Rev. {\bf B42}, 8145 (1990);
J. Fr\"ohlich and A. Zee, Nucl. Phys. {\bf  B364}, 517 (1991).
} 
\bibitem{wen1}{X.-G. Wen, Phys. Rev. B {\bf 43}, 11025 (1991); Phys. Rev.
Lett. {\bf 64}, 2206 (1990). }
\bibitem{haldane1}{F. D. M. Haldane, Phys. Rev. Lett. {\bf 47}, 1840 (1981);
J. M. Luttinger, J. Math. Phys. {\bf 4}, 1154 (1963); S. Tomonaga,
Prog. Theor. Phys. {\bf 5}, 544 (1950).}
\bibitem{beenakker}{C. W. J. Beenakker and H. van Houten, in {\it
Solid State Physics}, eds. H. Ehrenreich and D. Turnbull (Academic, New York,
1991), vol. 44.}
\bibitem{kfp}{C. L. Kane, M. P. A. Fisher, and J. Polchinski,
Phys. Rev. Lett. {\bf 72}, 4129 (1994).}
\bibitem{haldane3}{F. D. M. Haldane, Phys. Rev. Lett. {\bf 74}, 2090 (1995).}
\bibitem{kane3}{C. L. Kane and M. P. A. Fisher, Phys. Rev. B {\bf 52},
17393 (1995).}
\bibitem{buttiker}{M. B\"uttiker, Phys. Rev. B {\bf 38}, 9375 (1988).}
\bibitem{alphenaar}{B. W. Alphenaar, P. L. McEuen, R. G. Wheeler, and
R. N. Sacks, Phys. Rev. Lett. {\bf 64}, 677 (1990).}
\bibitem{kouwenhoven}{L. P. Kouwenhoven {\it et al.}, Phys Rev. Lett.
{\bf 64}, 685 (1990).}
\bibitem{takagaki}{Y. Takagaki {\it et al.}, Phys. Rev. B {\bf 50},
4456 (1994).}
\bibitem{longrange}{KFP argue that long-range Coulomb interactions included
in the low-energy field theory give corrections of order $\log^{-2}(L/\ell)$
to the conductance, with $L$ the screening length and $\ell$ the magnetic
length, and that such corrections are not observed experimentally.  Another
model is that described in the text: contact interactions but with
the charge-charge term larger than the charge-neutral term after
integrating the interaction up to the screening length.  In any event it
seems quite plausible that current experimental setups are almost
charge-unmixed.}
\bibitem{read}{N. Read, Phys. Rev. Lett. {\bf 65}, 1502 (1990).}
\bibitem{kane}{C. L. Kane and M. P. A. Fisher,
Phys. Rev. B {\bf 51}, 13449 (1994).}
\bibitem{haldane2}{F. D. M. Haldane, Phys. Rev. Lett. {\bf 51}, 605 (1983).}
\bibitem{halperin2}{B. I. Halperin, Phys. Rev. Lett. {\bf 52}, 1583 (1984).}
\bibitem{watson}{G. L. Watson, {\it Integral Quadratic Forms}
(Cambridge University Press, Cambridge, 1960), chaps. 1-3.}
\bibitem{giamarchi}{T. Giamarchi and H. J. Schultz, Phys. Rev. B
{\bf 37}, 325 (1988).}
\bibitem{misner}{C. W. Misner, K. S. Thorne and J. A. Wheeler,
{\it Gravitation} (W. H. Freeman and Co., New York, 1973), p. 69.}
\bibitem{eisenstein}{J. P. Eisenstein, H. L. Stormer, L. N. Pfeiffer, and
K. W. West, Phys. Rev. Lett. {\bf 62}, 1540 (1989); Phys. Rev. B {\bf 41},
7910 (1990).}
\bibitem{halperin3}{B. I. Halperin, Helv. Phys. Acta. {\bf 56}, 75 (1983).}
\bibitem{moore}{J. E. Moore and F. D. M. Haldane, Phys. Rev. B
{\bf 55}, 7818 (1997).}
\bibitem{johnson}{M. D. Johnson and A. H. MacDonald, Phys. Rev. Lett.
{\bf 67}, 2060 (1991).}
\bibitem{wenrev1}{X.-G. Wen, Int. J. Mod. Phys. B {\bf 6}, 1711 (1992).}
\bibitem{mcdonald}{I. A. McDonald and F. D. M. Haldane, Phys. Rev. B
{\bf 53}, 15845 (1996).}
\bibitem{moon}{K. Moon {\it et al.}, Phys. Rev. Lett. {\bf 71}, 4381
(1993).}
\bibitem{kane2}{C. L. Kane and M. P. A. Fisher, Phys. Rev. B {\bf 46},
15233 (1992).}
\end{references}
\end{document}